# Controlling Mixed Mo/$MoS_2$ Domains on Si by Molecular Beam Epitaxy for the Hydrogen Evolution Reaction

Eunseo Jeon[1,*], Vincent Masika Peheliwa[2,3,*], Marie Hrůzová Kratochvílová[2], Tim Verhagen[2,†], Yong-Kul Lee[1,†]

[1]Laboratory of Advanced Catalysis for Energy and Environment, Department of Chemical Engineering, Dankook University, Yongin 16890, South Korea

[2]Institute of Physics of the Czech Academy of Sciences, Prague 182 00, Czech Republic

[3]Faculty of Mathematics and Physics, Charles University, Prague 121 16., Czech Republic

* Equally contributed,

† To whom all correspondence should be addressed: verhagen@fzu.cz; yolee@dankook.ac.kr

**Abstract Graphic**

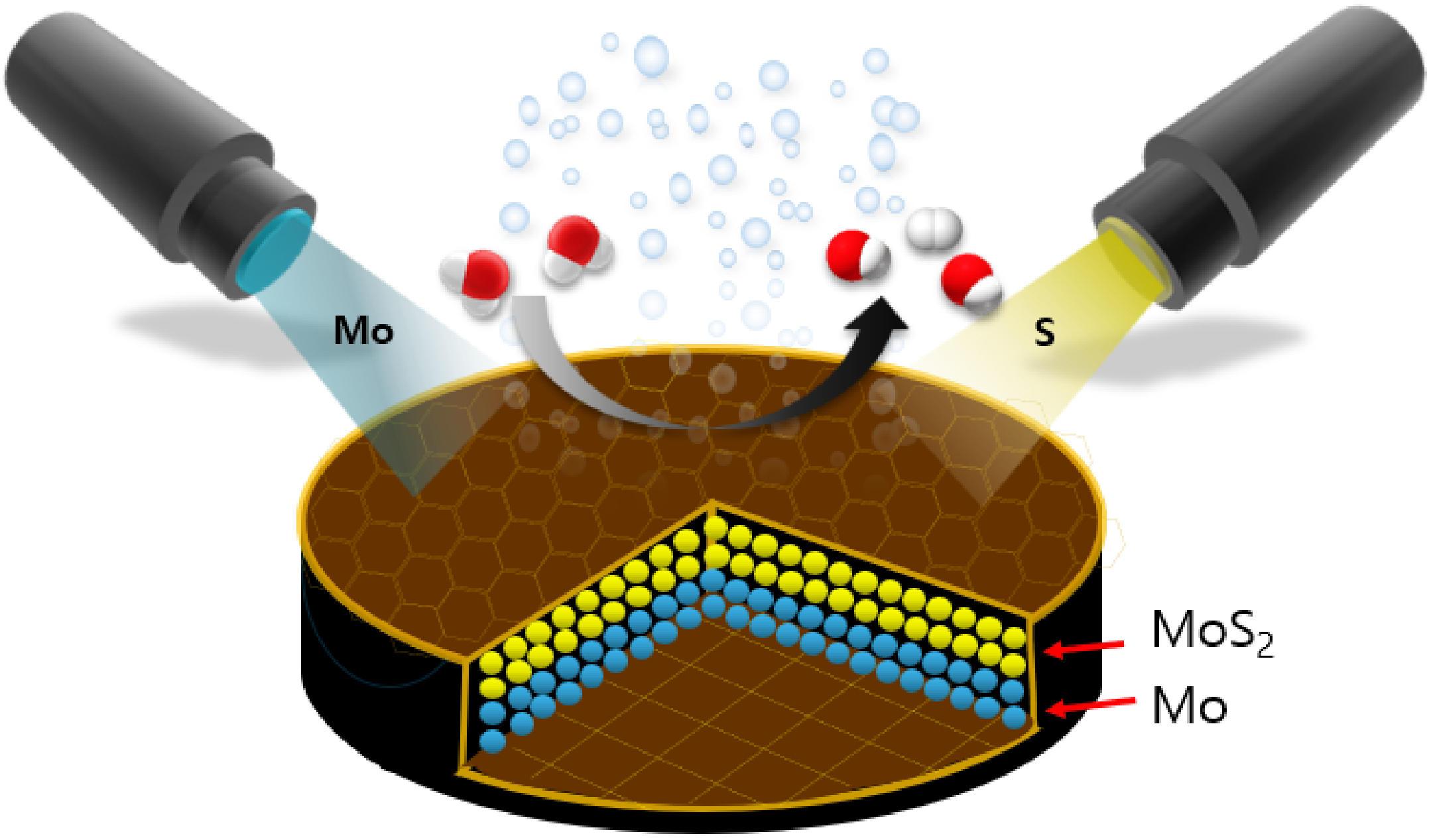

**Abstract**

Molybdenum disulfide ($MoS_2$) is a prototypical layered transition-metal dichalcogenide whose electrocatalytic performance is governed by a delicate balance between crystallinity, defect density, and electronic conductivity. Here we report a systematic molecular beam epitaxy (MBE) study in which annealing temperature, deposition cycle number, and Mo/S thickness ratio were independently varied to control the structural and electronic properties of $MoS_2$ thin films. The successful epitaxial growth of atomically uniform $MoS_2$ directly on Si substrates enables strong interfacial coupling and efficient charge transfer, offering a viable route toward semiconductor-integrated catalytic architectures. X-ray diffraction, Raman spectroscopy, and X-ray absorption analyses reveal that higher annealing temperatures and excessive deposition cycles enhance crystallinity but reduce edge-site density and electrical conductivity, leading to diminished hydrogen evolution reaction (HER) activity. In contrast, intermediate cycle numbers and sulfur-deficient growth conditions yield heterostructures composed of $MoS_2$ with residual metallic Mo and sulfur vacancies, which activate otherwise inert basal planes while providing conductive pathways. These defect-engineered films deliver the best catalytic performance, achieving overpotentials as low as −0.33 V at −10 mA $cm^{-2}$, enlarged electrochemical surface area (ECSA) up to 8.0 $cm^2$, and mass-based turnover frequencies exceeding 23 mmol $H_2$ $g^{-1}$ $s^{-1}$, more than double those of stoichiometric counterparts. Our findings establish sulfur stoichiometry and growth kinetics as powerful levers to tune the interplay between structural order and catalytic activity in MBE-grown $MoS_2$ and point toward a broader strategy for engineering layered catalysts at the atomic scale.

Electrochemical energy conversion and storage systems, including water electrolyzers, metal–air batteries, and solid-state batteries, are pivotal for the sustainable energy transition and future energy security.[1-3] Among diverse material candidates, transition metal chalcogenides (TMCs), and molybdenum disulfide ($MoS_2$) in particular, have become central due to earth abundance, tunable electronic structures, and remarkable catalytic activity for hydrogen evolution and other key reactions.[4-6] Early foundational studies established that the catalytic efficiency of $MoS_2$ originates from edge sites, providing a rationale for atomic-scale structure engineering.[1] Subsequent work revealed that sulfur vacancies and Frenkel defects on basal planes strongly influence intrinsic catalytic properties, driving systematic defect-engineering strategies.[3, 7] Recent advances in defect and interface engineering have shown that sulfur vacancies, local coordination distortions, and boundary restructuring play central roles in tuning the electronic and catalytic properties of $MoS_2$.[8, 9] Edge-oriented strategies, including deliberate nano-folding and boundary reconstruction, have further demonstrated substantial hydrogen evolution reaction (HER) enhancement by increasing the density of active edge terminations.[10] In parallel, post-growth defect patterning approaches have enabled precise modulation of $MoS_2$ transport and catalytic behavior through controlled sulfur-vacancy engineering.[11]

Progress has also been marked by the realization and stabilization of metallic 1T phases of $MoS_2$, which yield improved charge transport and activation of previously inert basal sites.[2] Boundary activation on monolayer $MoS_2$, strain engineering, and tuning allotrope-dependent activity–stability relationships further expanded the design space for catalytic optimization.[6, 12, 13] Hierarchical and hybrid architectures, including $MoS_2/CoSe_2$ nanobelts, porous foams, and heterostructured supports, have boosted accessible surface areas, enhanced mass and charge transport, and facilitated scalability.[14, 15] Parallel efforts in atomically engineered active sites, single-atom doping, and ensemble nanozyme concepts have underscored the importance of precision control in advancing catalytic performance,[5, 6] in which *in situ* spectroscopic and computational studies further highlighted mechanistic diversity, edge-basal interplay, and dynamic structural evolution under realistic operating conditions.

Despite these advances, precise and deterministic control over defects, stoichiometry, and structural ordering in $MoS_2$ remains a persistent challenge. Collectively, these studies establish conductivity, sulfur-vacancy density, and edge-site exposure as key descriptors of HER activity; however, how these parameters can be systematically and independently tuned and correlated within a single, well-defined growth platform remains fundamentally constrained by existing synthesis methods. In CVD or solution-based syntheses, Mo is typically fully sulfurized under sulfur-rich conditions, or Mo-rich phases such as $Mo_2N$, $Mo_2C$, or other $Mo_xSy$ compounds form as separate particles or heterostructured components, and these methods generally lack sufficiently precise control over crystallinity, defect density, and electronic transport, making it difficult to understand their individual contributions to catalytic performance.[16-21] As a result, the fundamental relationships between growth parameters, structural evolution, and catalytic activity remain incompletely understood.[2, 12, 18-21] Within this context, the deliberate formation of spatially coexisting metallic Mo and $MoS_2$ domains within a continuous thin film on Si is extremely challenging and, to the best of our knowledge, has not been systematically reported for HER electrocatalysis.

Molecular beam epitaxy (MBE) addresses these limitations by enabling systematic control of growth kinetics, sulfur stoichiometry, and post-growth annealing conditions.[22, 23] Recent advances have shown that MBE can yield wafer-scale, high-quality 2D films with controllable layer number and stoichiometry,[24-26] providing a means to systematically probe structure-property relationships. In this context, MBE growth directly on Si introduces a distinct, semiconductor-integrated regime, in which mixed Mo/$MoS_2$ domains, sulfur off-stoichiometry, and stacking order evolve on a non-catalytic support and collectively govern HER kinetics. This platform therefore complements exfoliation- and CVD-based approaches by providing a wafer-scale model system where growth parameters, interfacial structure, and electrochemical response can be probed within a single, well-defined architecture.

Here, we address this gap by systematically investigating the effects of annealing temperature, deposition cycle number, and sulfur flux on the structural and electrochemical properties of MBE-grown $MoS_2$ thin films. Using a combination of X-ray diffraction, Raman spectroscopy, X-ray absorption spectroscopy, and electrochemical analysis, we establish how crystallinity, sulfur stoichiometry, mixed Mo/$MoS_2$ domains, and stacking order evolve under different growth conditions and how these factors correlate with hydrogen evolution activity. To avoid post-hoc correlation, we explore a pre-defined growth parameter space in which only one parameter, annealing temperature, deposition cycle number, or sulfur flux, is varied at a time while all other conditions are kept constant.

## Results and discussion

To reduce the complexity of the $MoS_2$/Si system, we designed three systematic series of samples in which annealing temperature, deposition cycle number, and sulfur flux were varied one at a time, while all other growth parameters and the electrochemical protocol were kept identical. Each sample in these series was characterized using the same set of structural [X-ray diffraction (XRD), reflection high-energy electron diffraction (RHEED), atomic force microscopy (AFM), scanning transmission electron microscopy (STEM), X-ray absorption near-edge structure (XANES) / extended X-ray absorption fine structure (EXAFS)] and electrochemical [linear sweep voltammetry (LSV), Tafel, electrochemical impedance spectroscopy (EIS), electrochemical surface area (ECSA)]) measurements, enabling one-to-one comparison and minimizing post-hoc interpretation.

*Annealing temperature effect*

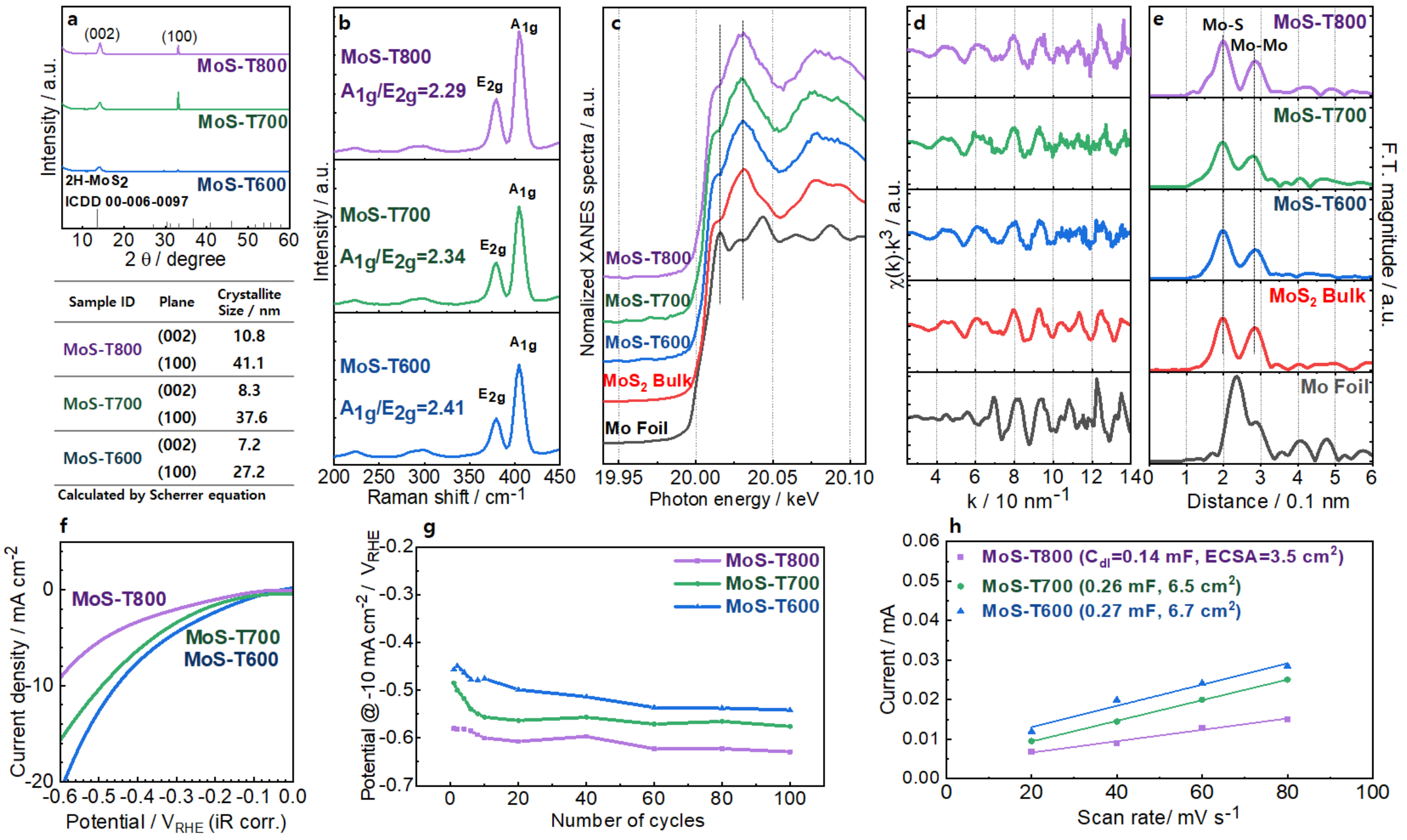

Figure 1. **a-e** Structural evolution of MBE-grown $MoS_2$ as a function of annealing temperature: **a** XRD patterns of $MoS_2$ thin films annealed at 600, 700, and 800 °C, showing sharpening of the (002) and (100) reflections indicative of enhanced crystallinity. **b** Raman spectra highlighting systematic changes in the $E_{2g}$ and $A_{1g}$ vibrational modes, with the decreasing $A_{1g}/E_{2g}$ intensity ratio reflecting reduced edge-site density at higher annealing temperatures. **c** Mo K-edge XANES spectra confirming the preservation of the 2H-$MoS_2$

phase across all annealed samples, characterized by the white-line feature at ~20,030 eV and shoulder at ~20,010 eV. **d, e** Corresponding EXAFS oscillations and Fourier transforms revealing Mo–S and Mo–Mo coordination shells (~0.197 nm and ~0.285 nm), consistent with bulk-like $MoS_2$ bonding environments. **f-h** Electrochemical performance of MBE-grown $MoS_2$ as a function of annealing temperature: **f** Polarization curves for $MoS_2$ films annealed at 600, 700, and 800 °C, showing that MoS-T600 exhibits the lowest overpotential (−0.456 V at −10 mA $cm^{-2}$), while higher annealing temperatures lead to diminished activity due to reduced edge-site density. **g** Long-term cycling stability over 100 cycles, demonstrating that all samples maintain robust activity without significant degradation under alkaline HER conditions. **h** Double-layer capacitance ($C_{dl}$) plots used to extract ECSA, revealing decreasing active-site density with increasing annealing temperature (6.7 $cm^2$ for T600, 6.5 $cm^2$ for T700, and 3.5 $cm^2$ for T800).

The structural evolution of molecular beam epitaxy (MBE)-grown $MoS_2$ with annealing temperature was investigated using XRD, Raman, and X-ray absorption fine-structure (XAFS) spectroscopy. Fig. 1a presents the diffraction patterns and crystallite sizes of $MoS_2$ thin films annealed at 600, 700, and 800 °C. The diffraction peaks indexed to the (002) and (100) planes of 2H-$MoS_2$ (JCPDS 00-006-0097) become sharper at elevated annealing temperatures, indicating enhanced crystallinity. Crystallite sizes, estimated using the Scherrer equation, increased from 7.2 nm to 10.8 nm for the (002) plane and from 27.2 nm to 41.1 nm for the (100) plane as the annealing temperature rose from 600 °C to 800 °C. X-ray reflectivity (XRR) profiles revealed reduced Kiessig fringe amplitude and smoother decay, indicating improved film density and surface smoothness as the angle where fringes are no longer resolvable increases up to ~4°with higher temperature (Fig S1a). Correspondingly, the differential reflectance spectra exhibited sharper A (∼1.9 eV) and B (∼2.1 eV) excitonic peaks, confirming enhanced crystallinity and reduced defect scattering in the annealed $MoS_2$ layers (Fig. S1b). RHEED patterns of MoS-T600, MoS-T700, and MoS-T800 showed sharper and more continuous rings with increasing annealing temperature, indicating enhanced crystallinity and surface ordering (Fig. S2). The corresponding intensity profiles (Fig. S2d) display peaks indexed to 2H-$MoS_2$ with an in-plane lattice constant of a = 3.164 Å, confirming improved structural quality after annealing. AFM images of MoS-T600, MoS-T700, and MoS-T800 showed that surface roughness decreases with increasing annealing temperature (Fig. S3). MoS-T600 exhibits granular features (root mean square roughness $R_{(q)}$ = 4.81 ± 0.81 nm), MoS-T700 shows coalesced grains ($R_{(q)}$ = 5.56 ± 1.10 nm), and MoS-T800 forms a smoother, compact surface ($R_{(q)}$ = 1.78 ± 0.35 nm), indicating enhanced crystallization at higher temperatures. These results confirm that post-growth annealing promotes ordering along both the basal and lateral directions in MBE-deposited films, consistent with prior observations of recrystallization in MBE-grown $MoS_2$ monolayers.[27] Raman spectroscopy provided further insight into the vibrational properties of the annealed $MoS_2$ films (Fig. 1b). Two prominent peaks were observed, corresponding to the in-plane $E_{2g}$ mode and the out-of-plane $A_{1g}$ mode. The $A_{1g}/E_{2g}$ intensity ratio decreased from 2.41 (600 °C) to 2.29 (800 °C), suggesting a reduction in edge-site density associated with the development of the horizontal (100) plane. This shift is consistent with a measurable reduction in edge-site density and increased grain coalescence at higher annealing temperature, as evidenced by the concurrent decrease in ECSA (from 6.7 to 3.5 $cm^2$) and the increase in crystallite size extracted from XRD analysis. Moreover, these results are supported by prior density functional theory (DFT) studies showing that vibrational responses of $MoS_2$ are highly sensitive to edge exposure and local defect environments: Li *et al.* demonstrated that the loss of edge terminations leads to a measurable reduction in Raman edge-related contributions,[28] and Komsa and Krasheninnikov showed that sulfur vacancies and edge-associated Mo d-states strongly modify Raman-active phonon modes through changes in electron–phonon coupling.[29] A fully quantitative theoretical conversion of the $A_{1g}/E_{2g}$ ratio into absolute edge length or active-site density would require Raman tensor calculations for specific edge geometries, which is beyond the intended scope of this work, but the observed trend aligns well with these established theoretical insights. In addition to the dominant $E_{2g}$ and $A_{1g}$ modes of 2H-$MoS_2$, weak features can originate from defect-activated LA(M) and 2LA(M) modes, which become Raman-allowed when sulfur vacancies or edge sites relax momentum selection rules.[29] In our samples, these modes remain weak, indicating a moderate defect density rather than strong lattice disorder. Similar trends in vibrational sharpening and defect suppression have been reported for annealed MBE-grown TMD layers.[27] The local structural and electronic properties of MBE-grown $MoS_2$ films annealed at 600, 700, and 800 °C were examined by Mo K-edge XANES and EXAFS spectroscopy. As shown in Fig. 1c, the XANES spectra of the annealed films exhibit a pronounced white-line feature at ~20,030 eV and a shoulder at ~20,010 eV, both characteristic of bulk 2H-$MoS_2$. These features confirm the preservation of the $MoS_2$ phase after annealing. The corresponding EXAFS spectra and Fourier transforms (FT) further highlight the well-defined local structure of $MoS_2$ (Fig. 1d and 1e). The FT magnitude spectra reveal distinct coordination shells at 0.197 nm (Mo–S) and 0.285 nm (Mo–Mo), consistent with the expected nearest-neighbor distances in crystalline 2H-$MoS_2$. Importantly, the MoS-T600, MoS-T700, and MoS-T800 samples all display spectral features closely matching those of bulk $MoS_2$, while remaining distinct from the metallic Mo reference foil. These observations indicate that MBE-grown $MoS_2$ films retain the intrinsic structural characteristics of bulk $MoS_2$ across the investigated annealing range. Such stabilization of the layered phase is consistent with earlier MBE studies showing robust epitaxial growth of highly crystalline $MoS_2$ and $MoSe_2$ on insulating substrates.[30]

The catalytic performance of MBE-grown $MoS_2$ films annealed at 600, 700, and 800 °C was evaluated for the hydrogen evolution reaction (HER) in alkaline electrolyte. As shown in Fig. 1f-h, the polarization curves of the first cycle reveal that MoS-T600 exhibits the highest activity, achieving an overpotential of −0.456 $V_{RHE}$ (vs. reversible hydrogen electrode (RHE), iR-corrected) at a current density of −10 mA $cm^{-2}$. In contrast, MoS-T700 and MoS-T800 display higher overpotentials, consistent with the reduced density of active sites as annealing temperature increases (Fig. 1f). Long-term cycling stability tests (up to 100 cycles) demonstrate that all annealed samples maintain stable catalytic activity without significant performance degradation, indicating robust structural stability under HER operating conditions (Fig. 1g). ECSA, determined from the double-layer capacitance ($C_{dl}$), further supports the observed activity trend. The ECSA values were estimated as 6.7 cm² ($MoS_2$-T600), 6.5 cm² ($MoS_2$-T700), and 3.5 cm² ($MoS_2$-T800), reflecting the reduction in accessible active sites with increasing crystallite size (Fig. 1h). This trend is consistent with the structural analysis, which showed enhanced crystallinity but fewer edge sites at higher annealing temperatures. Moreover, resistivity systematically increases with annealing temperature: 15.98 Ω·cm (T600), 16.52 Ω·cm (T700), and 19.26 Ω·cm (T800), as summarized in Table 1. This trend indicates that while higher annealing temperatures enhance crystallinity and enlarge lateral crystallite size, they also reduce the density of catalytically active edge sites. The reduction in structural disorder and defects decreases carrier density and available conductive pathways, thereby increasing bulk resistivity. EIS provides quantitative evidence for how annealing-driven structural evolution impacts electron transport during HER. While the bulk resistivity difference between 600 °C and 700 °C is small, the charge-transfer resistance ($R_{ct}$) extracted from Nyquist fitting increases with annealing temperature and becomes substantially larger for the 800 °C film, indicating that interfacial charge transfer is progressively hindered as defects and edge-related transport channels are reduced. This trend is consistent with the concomitant decrease in ECSA and Raman signatures associated with reduced edge/defect density at higher annealing temperatures, supporting the interpretation that decreasing structural disorder limits defect-mediated conductive pathways and lowers the effective electron-transport efficiency during HER. Thus, although high-temperature annealing improves structural ordering, it has the adverse effect of suppressing electronic conductivity, which directly contributes to the observed decline in HER activity. These findings align with previous reports on heteroepitaxial MBE growth of $MoS_2$/graphene and $MoS_2$/h-BN systems, which highlight the trade-off between crystallinity and conductivity.[31, 32] Overall, these results highlight the critical balance between crystallinity and catalytic activity in $MoS_2$. While high-temperature annealing improves the structural ordering of MBE-grown films, it reduces edge-site density and thereby diminishes HER performance. MoS-T600, with its relatively small crystallite size and high density of catalytically active edge sites, achieves the most favorable balance, delivering the best HER activity among the tested samples.

*Deposition cycle number effect*

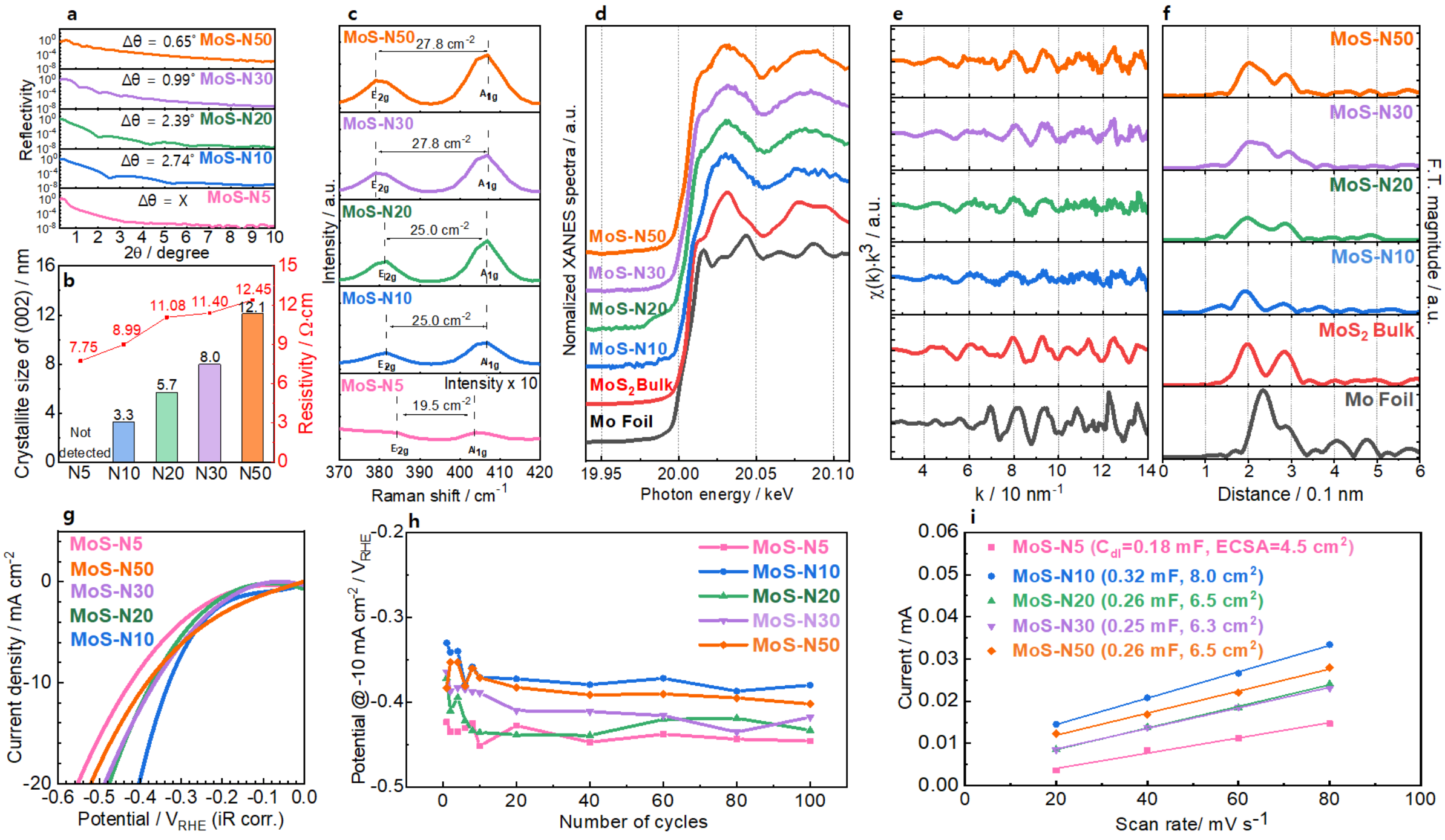

Figure 2. **a-f** Structural and electronic evolution of MBE-grown $MoS_2$ as a function of deposition cycle number: **a, b** XRR and resistivity analyses showing progressive thickening and smoother surfaces with increasing deposition cycles, accompanied by a monotonic rise in resistivity from N10 to N50. **c** Raman spectra of $MoS_2$ films, highlighting systematic shifts in the $E_{2g}$ (red-shift) and $A_{1g}$ (blue-shift) modes with increasing layer number, consistent with stronger interlayer interactions. **d** Mo K-edge XANES spectra confirming the preservation of the $MoS_2$ phase across all deposition cycles, with spectral features characteristic of bulk 2H-$MoS_2$. **e, f** Corresponding EXAFS oscillations and Fourier transforms showing intensification of Mo–S and Mo–Mo coordination peaks as layer number increases, reflecting enhanced structural ordering and bulk-like bonding. **g-i** Electrocatalytic performance of MBE-grown $MoS_2$ as a function of deposition cycle number: **g** Polarization curves for $MoS_2$ films with different deposition cycles, showing that MoS-N10 exhibits the lowest overpotential (−0.326 V at −10 mA $cm^{-2}$), while both thinner (N5) and thicker (N30, N50) films display inferior activity. **h** Long-term cycling stability over 100 cycles, demonstrating that all samples retain stable activity under alkaline HER conditions without significant degradation. **i** Double-layer capacitance ($C_{dl}$) plots used to extract ECSA, revealing the largest accessible surface area for MoS-N10 (8.0 $cm^2$), in contrast to reduced values for higher cycle numbers (N30, N50).

The structural evolution of MBE-grown $MoS_2$ with deposition cycle number was investigated using XRR measurements, Raman, and XAFS spectroscopy (Fig. 2a-f). As the fringe thickness ($\Delta\theta$) is inversely proportional to the film thickness, XRR results confirm that the $MoS_2$ films become progressively thicker with increasing deposition cycles. (Fig. 2a) Simultaneously, the (002) diffraction peak intensity in XRD increases with deposition cycles (Fig. S6a), and the differential reflectance spectra exhibit stronger and sharper excitonic features near 1.9 and 2.1 eV with increasing cycle number (Fig. S6b) , reflecting improved structural ordering and enhanced growth along the basal plane direction, in line with prior observations of improved crystallinity *via* precise layer control in MBE.[33] Crystallite sizes based on the (002) peak increase significantly from 3.3 nm (N10) to 12.1 nm (N50), consistent with layer-by-layer thickening of the $MoS_2$ lattice. RHEED and AFM analysis (Figs. S7 and S8) revealed a progressive transition from amorphous to crystalline morphology as the number of deposition cycles increases. MoS-N5 exhibits no RHEED features, indicating poor crystallinity, whereas MoS-N10 to MoS-N30 show streak patterns characteristic of layered growth, and MoS-N50 displays ring-like features corresponding to polycrystalline domains. The corresponding intensity profiles confirm a gradual increase in long-range order and the in-plane lattice constant (~3.16 Å) consistent with 2H-$MoS_2$. The root mean square (RMS) roughness in AFM images remained nearly constant (0.47–0.75 nm) as the deposition cycle increased from 5 to 50 layers, indicating smooth and uniform film growth. Electrical measurements indicate a monotonic increase in resistivity with cycle number, ranging from 9.0 (N10) to 12.5 Ω·cm (N50), attributable to increased thickness and crystallite size that suppress vertical electron transport across van der Waals layers (Fig. 2b). This trade-off between structural quality and conductivity parallels findings in $MoS_2$/graphene heterostructures fabricated by MBE, where enhanced ordering came at the expense of charge transport across thicker or more ordered regions.[31] Raman spectroscopy further elucidates deposition-cycle effects: the in-plane $E_{2g}$ mode red-shifts (due to dielectric screening) and the out-of-plane $A_{1g}$ mode blue-shifts (due to interlayer vdW stiffening), with increasing $E_{2g}$–$A_{1g}$ separation, a known fingerprint of increased layer number.[27] (Fig. 2c) These spectral trends are similarly observed in heteroepitaxial $MoS_2$ films grown on h-BN by MBE.[27] XANES and EXAFS spectroscopy revealed that all samples maintain the fundamental features of 2H-$MoS_2$, including the distinct white-line peak near 20,030 eV and the shoulder around 20,010 eV, indicative of Mo–S coordination (Fig. 2d-e). Fourier-transformed EXAFS spectra display characteristic Mo–S (~0.197 nm) and Mo–Mo (~0.285 nm) coordination shells, confirming local structural ordering (Fig. 2f). A slight deviation in oscillation amplitude and phase compared with bulk $MoS_2$ suggests partial metallic Mo contributions or incomplete sulfidation during early-stage growth. These residual metallic domains are likely embedded at the $MoS_2$-substrate interface, consistent with the observed reduction in oscillation intensity. A more detailed investigation of the metallic phase evolution and its correlation with Mo/S stoichiometry is presented in the following section. Strengthening of EXAFS peaks with cycle number reflects improved ordering and coordination, reminiscent of similar structural stabilization in initial stages of MBE growth for $MoSe_2$,[30] and robust epitaxy observed in MBE-grown $MoS_2$ monolayers.[27]

As shown in Fig. 2g-i, catalytically, $MoS_2$-N10 attains the best HER performance (overpotential $\eta = -0.326$ V at $-10$ mA $cm^{-2}$), balancing thickness, surface accessibility, and conductivity. While even thicker films (N30–N50) are structurally superior, their diminished conductivity reduces HER activity, demonstrating a clear trade-off also highlighted in previous studies of MBE-grown layered films.[31, 34] N10 offers the optimal combination of ECSA and charge transfer, consistent with findings that intermediate thicknesses in MBE-grown $MoS_2$ optimize electronic and catalytic behavior.

*Monolayer thickness effect*

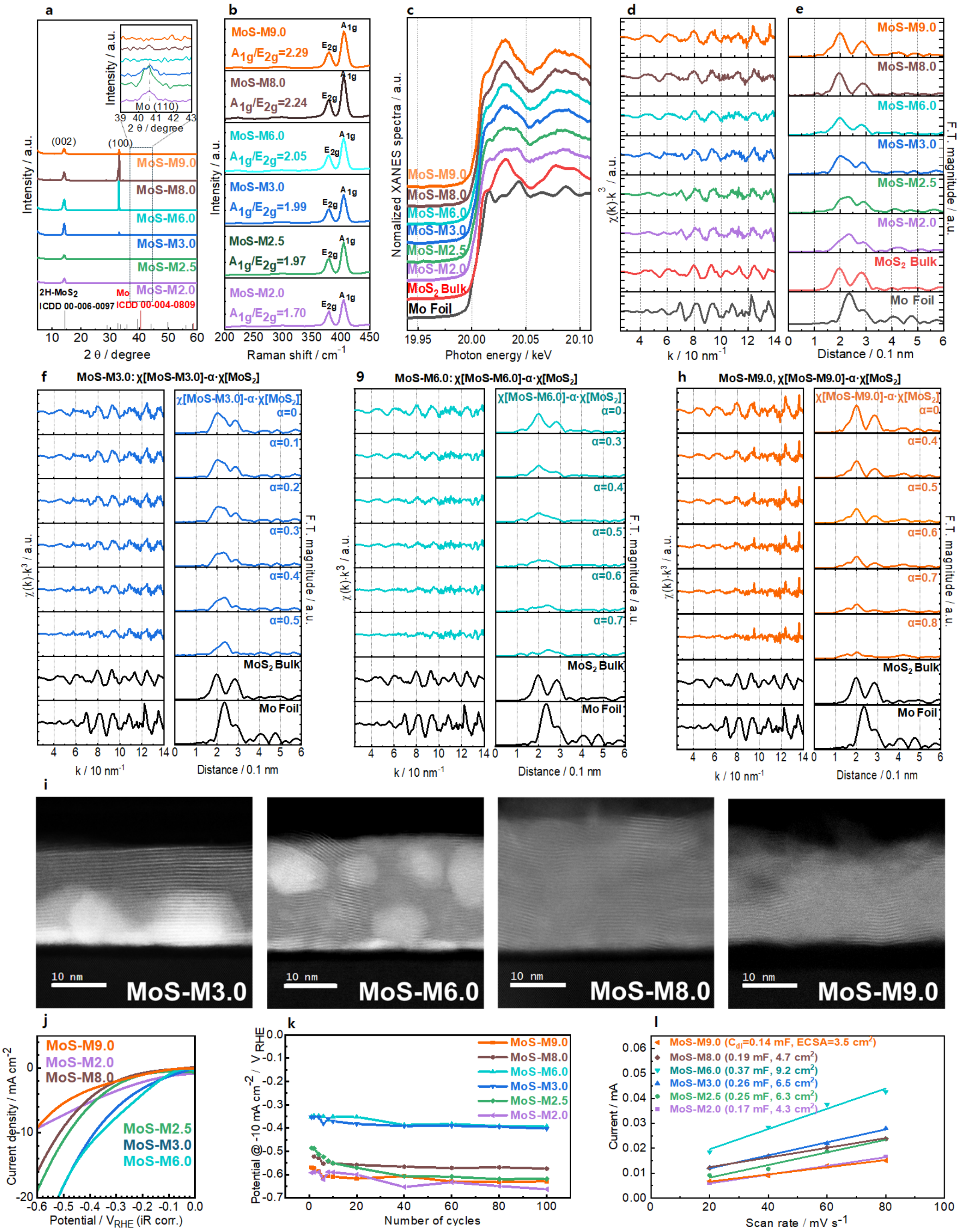

Figure 3. **a-e** Structural and electronic evolution of MBE-grown $MoS_2$ as a function of sulfur supply: **a** XRD patterns of $MoS_2$ films grown with varying sulfur thickness (2.0–9.0), showing incomplete sulfurization and residual metallic Mo reflections at low sulfur supply (M2.0–M3.0), and the exclusive presence of 2H-$MoS_2$ reflections at higher sulfur levels (≥ M6.0). **b** Raman spectra highlighting systematic changes in the $A_{1g}/E_{2g}$ intensity ratio, which increases with sulfur supply (1.70 for M2.0 to 2.29 for M9.0), reflecting enhanced sulfur termination and out-of-plane vibrational contributions. **c** Mo K-edge XANES spectra showing the transition from mixed-phase (metallic Mo + $MoS_2$) at low sulfur supply to pure $MoS_2$ at higher sulfur flux, with the emergence of characteristic white-line and shoulder features of bulk 2H-$MoS_2$. **d, e** Corresponding EXAFS oscillations and Fourier transforms, where metallic Mo–Mo contributions diminish with increasing sulfur thickness, and Mo–S (~0.197 nm) and Mo–Mo (~0.285 nm) coordination shells characteristic of bulk $MoS_2$ become dominant. **f-g** Local bonding environment and microstructural evolution of MBE-grown $MoS_2$ as a function of sulfur supply: **f** EXAFS subtraction analysis for MoS-M3.0, showing residual metallic Mo contributions coexisting with $MoS_2$ coordination shells, indicative of incomplete sulfurization. **g** EXAFS subtraction analysis for MoS-M6.0, where metallic Mo features are partially suppressed and Mo–S coordination becomes dominant, reflecting intermediate sulfurization. **h** EXAFS subtraction analysis for MoS-M9.0, exhibiting spectra that closely resemble bulk $MoS_2$, confirming near-complete sulfurization under high sulfur supply. **i** STEM images of $MoS_2$ films grown at different sulfur thicknesses, illustrating the microstructural transition from (i) layered $MoS_2$ stacked on metallic Mo (M3.0), to (ii) intercalated metallic Mo domains within the layered structure (M6.0), to (iii) well-aligned $MoS_2$ layers at near-optimal stoichiometry (M8.0), and finally to (iv) partially disordered stacking induced by excessive sulfur supply (M9.0). **j-l** Electrocatalytic performance of MBE-grown $MoS_2$ as a function of sulfur supply: **j** Polarization curves of $MoS_2$ films prepared with different sulfur thicknesses, showing that S-deficient samples (M2.5, M3.0, M6.0) deliver superior activity compared with sulfur-rich films (M8.0, M9.0), with MoS-M3.0 and M6.0 achieving the lowest overpotential. **k** Cycling stability tests over 100 cycles, demonstrating robust structural and catalytic stability for all samples under alkaline HER conditions. **l** Double-layer capacitance ($C_{dl}$) measurements used to estimate ECSA, revealing larger accessible areas for S-deficient films (up to 9.2 $cm^2$ for M6.0) relative to sulfur-rich samples (3–5 $cm^2$).

The effect of sulfur supply on phase formation and structural ordering in MBE-grown $MoS_2$ was systematically investigated by varying the sulfur deposition thickness from 2.0 to 9.0 Å per Mo layer (~3 Å), denoted as M2.0-M9.0. XRD analysis revealed that samples grown under insufficient sulfur flux (MoS-M2.0, M2.5, M3.0) retained metallic Mo peaks, indicating incomplete sulfurization (Fig. 3a). In contrast, higher sulfur deposition (≥ 6.0) yielded clear 2H-$MoS_2$ phases, consistent with earlier chemical vapor deposition (CVD) and chemical vapor sulfurization studies showing that sulfur flux critically dictates phase purity and crystallization pathways.[35, 36] Raman spectroscopy confirmed this trend: $E_{2g}$ and $A_{1g}$ modes appeared in all films, with the $A_{1g}/E_{2g}$ intensity ratio rising from 1.70 (M2.0) to 2.29 (M9.0) (Fig. 3b). This reflects increased sulfur-terminated edges and out-of-plane vibrations at higher sulfur supply. Notably, the absence of LA(M) or 2LA(M) Raman modes indicates that the defect density remains moderate, suggesting that the observed morphological variations arise mainly from stoichiometric imbalance rather than defect-induced lattice distortion. The ratio is further influenced by the relative fraction of grains oriented along the (100) versus (002) planes, as evident from the STEM images and corroborated by Raman data. Similar Raman shifts due to sulfur overpressure and defect density modulation have been documented in sulfur-vacancy studies,[37] as well as in atomically etched $MoS_2$ where defect-driven vibrational changes were highlighted.[38] The XRR profiles showed systematic changes in film density and interface quality with increasing sulfur deposition (Fig. S11a). At low sulfur supply (MoS-M2.0 to MoS-M3.0), clear oscillatory fringes are observed, indicating uniform layer growth. As the sulfur supply increases to MoS-M6.0, the reflectivity curve becomes smoother with distinct oscillation amplitude and decreased critical angle, consistent with reduced mass density and improved structural uniformity. However, further sulfur enrichment (MoS-M8.0 and MoS-M9.0) leads to dampened reflectivity and a loss of periodic features, suggesting surface and interface roughening and partial disorder formation, possibly due to sulfur over-saturation and nonstoichiometric phase segregation. Differential reflectance spectra reveal a gradual enhancement and sharpening of the A (∼1.9 eV) and B (∼2.1 eV) excitonic peaks up to MoS-M6.0, followed by peak broadening at higher sulfur doses (M8.0–M9.0), suggesting optimal crystallinity and minimal defect scattering at moderate sulfur stoichiometry (Fig. S11b). RHEED and AFM analysis collectively revealed that sulfur supply critically governs the phase formation and surface morphology of MBE-grown $MoS_2$ films (Figs S12 and S13). Low-sulfur samples (M2.0–M3.0) show diffuse RHEED features and granular AFM morphologies, indicating incomplete sulfidation and partial metallic Mo character. With increased sulfur flux (M6.0–M9.0), sharp RHEED rings corresponding to 2H-$MoS_2$ reflections (a = 3.164 Å) emerge, confirming improved crystallinity and phase purity. Concurrently, surface roughness decreases to a minimum ($R_a \approx$ 0.56 nm for M6.0), then slightly increases again under excessive sulfur, consistent with partial disorder and stacking defects. These results demonstrate that moderate sulfur supply yields the most uniform and crystalline $MoS_2$ layers, while both deficiency and excess induce structural imperfections. XANES spectra showed a clear transition from mixed-phase (Mo + $MoS_2$) at low sulfur to pure $MoS_2$ at high sulfur flux (Fig. 3c). EXAFS analysis supported this evolution: Mo–Mo metallic peaks diminished while Mo–S coordination dominated (Fig. 3d and e). These observations align with prior findings that sulfur vacancies strongly modulate local coordination environments and can act as catalytic centers.[16] Similar stabilization of Mo–S coordination shells in controlled vacancy-engineered $MoS_2$ has been reported.[3]

Because EXAFS averages over heterogeneous coordination environments, the mixed-phase ($MoS_2$ + metallic Mo) signals cannot be fully resolved by single-phase fitting alone. We therefore performed a stepwise subtraction of the bulk-$MoS_2$ spectrum, $\chi_{\mathrm{res}}(k) = \chi_{\mathrm{sample}}(k) - \alpha\chi_{\mathrm{MoS2,bulk}}(k)$ with $\alpha$ = 0.0-1.0, as shown in Fig. 3f-h. The intensity of the residual Mo–Mo path (≈0.285 nm; R-space peak before phase correction) serves as a proxy for the metallic component, revealing a monotonic decrease M3.0 > M6.0 $\gg$ M9.0. STEM images further visualized this progression (Fig. 3i). MoS-M3.0 exhibited $MoS_2$ layers atop metallic Mo, MoS-M6.0 revealed intercalated metallic domains, MoS-M8.0 showed well-aligned layered order, while MoS-M9.0 exhibited stacking disorder. This observation resonates with recent work where excessive sulfur introduced disorder in otherwise crystalline TMD lattices.[39]

In sulfur-rich $MoS_2$ films, diminished conductivity is manifested as a pronounced increase in charge-transfer resistance ($R_{ct}$) extracted from EIS (Table 1). This behavior indicates that the conductivity loss arises from interfacial and film-limited transport. Structurally, excess sulfur suppresses metallic Mo–Mo conductive pathways and reduces sulfur-vacancy–derived donor states, leading to a lower effective carrier density, while increased stacking order further restricts carrier mobility. Consistent with this interpretation, sulfur-rich samples exhibit higher $R_{ct}$, larger Tafel slopes, and higher resistivity together with reduced ECSA, demonstrating that impaired charge transport is the dominant cause of the diminished conductivity and HER activity. Catalytic HER performance strongly reflected these structural trends. Surprisingly, S-deficient films (M2.5, M3.0, M6.0) showed better activity than sulfur-rich films (M8.0, M9.0). The synergistic coexistence of metallic Mo and $MoS_2$ improved conductivity and provided abundant catalytic sites. This agrees with prior studies demonstrating that metallic domains enhance electron transfer while sulfur vacancies activate basal planes, otherwise inert in stoichiometric films.[36] Furthermore, computational and experimental analyses have shown that sulfur-deficient or mixed-phase catalysts often outperform fully stoichiometric analogues in HER.[40-42] ECSA measurements supported this: S-deficient samples had larger ECSA values (6.5-9.2 $cm^2$), whereas sulfur-rich samples exhibited smaller values (3-5 $cm^2$). This is consistent with reports that sulfur vacancies enlarge the accessible active surface area and provide basal-plane reactivity.[3, 43] These results demonstrate that while sulfur oversupply ensures phase purity and well-formed layered $MoS_2$, it compromises HER performance by eliminating conductive metallic domains and defect-induced active sites. In contrast, intermediate sulfur supply produces mixed Mo + $MoS_2$ phases that combine conductivity, defect activation, and edge reactivity. This finding mirrors a broader consensus in the literature that controlled imperfection *via* sulfur vacancies or partial sulfurization can yield superior electrocatalysts compared to perfectly stoichiometric crystals.[16, 39, 43] By tuning the sulfur flux during MBE growth, we systematically vary the metallic Mo/$MoS_2$ fraction, sulfur-vacancy density, and stacking order, which in turn produce consistent trends in $R_{ct}$, Tafel slopes, ECSA, and overpotential. Rather than isolating a single descriptor, the MBE-on-Si platform provides a controlled, reproducible structure–activity landscape where conductivity, defect chemistry, and edge accessibility co-evolve to govern HER performance. Importantly, these trends do not merely reproduce previously reported correlations between defects and HER activity in exfoliated or CVD-grown $MoS_2$, but instead elucidate how metallic Mo domains, sulfur-vacancy density, and stacking order co-operate within a Mo/$MoS_2$/Si architecture to govern charge transfer and active-site accessibility. In this context, the present work extends established HER descriptors into a semiconductor-integrated platform enabled by MBE growth, rather than proposing entirely new descriptors.

*Structure-growth correlation of $MoS_2$*

The epitaxial growth of well-aligned $MoS_2$ multilayers is inherently challenging due to the weak van der Waals interactions governing interlayer coupling. The variety of structural motifs observed in the present $MoS_2$ films (Figs. 1–3) clearly illustrates this complexity. STEM imaging of films with different sulfur stoichiometry provides key insights: in MoS-M3.0, where residual Mo particles are incorporated, the $MoS_2$ layers exhibit relatively ordered stacking along the (002) direction. In contrast, films with reduced Mo particle density display greater disorder, consistent with less effective screening of electrostatic boundary conditions. Similarly, thinner MoS-Nx films remain preferentially aligned along the (002) direction, while thicker films increasingly adopt a (100) orientation, reflecting a transition in stacking preference with increasing thickness.

In recent years, it has been shown that each nonpolar van der Waals material can be straightforwardly turned into a polar or even ferroelectric van der Waals material when a slide or twist is applied to the layer, thereby breaking the inversion and/or mirror symmetry, the so-called sliding ferroelectricity.[44-46] An important difference with conventional ferroelectric materials, where the nonzero polarization is induced by ionic displacement, is that the polarization is solely created due to the stacking of the layers in van der Waals materials.

If the structural effects observed here are compared with thin polar or ferroelectric oxide films grown and annealed in ultra-high vacuum by MBE or pulsed laser deposition (PLD), similarities can be observed to control the electrostatic boundary conditions.[47] The induced charge in the sliding ferroelectric $MoS_2$-stack can be screened by a metallic 'electrode', here the Mo particles, and $MoS_2$ layers are dominantly stacked in the (002)-direction. When less Mo particles are present, the depolarizing field can be reduced via the formation of in-plane domains. In sliding ferroelectrics, to rotate the out-of-plane polarization direction into an in-plane polarization direction, the $MoS_2$ layer stacking should be re-oriented from the (002) to the (100)-orientation, as can be seen in the STEM figures in Fig. 3i. Furthermore, off-stoichiometry can contribute to the reduction of the depolarization field, which is here the presence of sulfur defects. Finally, as demonstrated in exfoliated 3R-stacked $MoS_2$, polarization saturates after approximately 10 layers due to polarization-induced bandgap closure.[48] Consistently, in the present work, a marked structural transition occurs as the thickness increases from 10 to 20 layers, reinforcing the strong interplay between thickness, stacking order, and polarization phenomena in MBE-grown $MoS_2$.

While mixed Mo/$MoS_2$ domains are spectroscopically evidenced by Mo–Mo coordination in EXAFS and systematic XANES edge shifts, their role in HER enhancement should be interpreted cautiously. In the present study, metallic Mo is not invoked as a dominant or isolated active phase, but rather as part of a mixed electronic landscape that is consistent with reduced charge-transfer resistance and improved interfacial electronic coupling observed in EIS. Similarly, possible effects related to interlayer sliding and polarization are discussed only as secondary, literature-informed interpretations and are not experimentally demonstrated here. Importantly, the primary conclusions of this work rely on directly supported correlations among sulfur stoichiometry, defect density, stacking order, charge-transfer resistance, and Tafel kinetics, which collectively govern the observed HER performance trends. Moreover, it is important to note that AFM roughness and ECSA probe fundamentally different aspects of surface structure and are therefore not expected to correlate directly in ultrathin $MoS_2$ films. AFM roughness reflects mesoscale height variations related to grain size and surface coalescence, whereas ECSA is governed by the density of electrochemically accessible active sites, which in $MoS_2$ are dominated by atomic-scale edges and defects beyond AFM resolution. Consequently, annealing-induced surface smoothing can coincide with reduced ECSA due to edge-site loss, while moderately defective films may exhibit low roughness yet high ECSA owing to enhanced edge accessibility. Accordingly, we do not claim atomically precise decoupling of individual descriptors, but rather a consistent, MBE-controlled correlation between structural motifs and electrochemical response across a well-defined parameter space. Instead, this decoupling is well established in layered TMDC systems: Jaramillo *et al.* showed that HER activity in $MoS_2$ scales with edge-site density rather than surface roughness,[1] and Voiry *et al.* further demonstrated that defect-mediated conductive pathways dominate HER performance even in morphologically smooth films.[49] These established insights are fully consistent with the trends observed in this study.

*Structure-activity correlation of $MoS_2$ in HER*

Table 1 summarizes the structural parameters, growth conditions, and electrochemical performance of MBE-grown $MoS_2$ thin films, while Fig. 4 visualizes the correlations between deposition cycles or sulfur stoichiometry and catalytic activities (|η|, ECSA or $MoS_2$ mass-based turnover frequency TOF). The comparative analysis summarized in Table 1 reveals that, while annealing temperature primarily affects crystallinity and edge-site density, its overall impact on HER activity is relatively modest. Across $MoS_2$-T600 to T800, the overpotential only changes from -0.46 to -0.58 V, with ECSA decreasing from 6.7 to 3.5 $cm^2$ and TOF (ECSA-based) remaining nearly constant at around ~5.7 nmol $H_2$ $cm^{-2}$ $s^{-1}$. This suggests that annealing-induced improvements in structural ordering are offset by the loss of active edge sites, leading to limited performance enhancement. By contrast, deposition cycle number exerts a much stronger influence. MoS-N10 exhibits the most favorable balance of conductivity and surface accessibility, achieving the lowest overpotential (−0.33 V), the largest ECSA (8.0 $cm^2$), and the highest ECSA-based TOF of 13.0 nmol $H_2$ $cm^{-2}$ $s^{-1}$. Remarkably, its mass-based TOF reaches 24.9 mmol $H_2$ $g^{-1}$ $s^{-1}$, demonstrating superior intrinsic activity normalized by catalyst loading (Table S1). Both insufficient cycles (N5) and excessive cycles (N20–N50) result in inferior performance, underscoring the critical role of thickness optimization. Interestingly, the enhanced charge transport in this intermediate-thickness regime may also stem from interlayer polarization effects caused by subtle sliding in van der Waals stacks, reminiscent of recently reported enhancement of HER in twisted $NbS_2$.[50] Such interlayer polarization could contribute to enhanced carrier localization, thereby facilitating charge transfer during the hydrogen evolution reaction. This localization of a high density of electronic states resembles the effects typically achieved through edge or defect engineering, but in this case extends across the entire twisted basal plane, as we have observed already for the misfit-layer compound $(PbS)_{1.11}VS_2$.[51] Similarly, sulfur stoichiometry during growth governs phase purity and active-site density (Fig. 4a). Sulfur-deficient films (M2.5-M6.0) outperform stoichiometric or sulfur-rich films, with $MoS_2$-M6.0 delivering an overpotential of -0.35 V, ECSA of 9.2 $cm^2$, an ECSA-based TOF of 6.7 nmol $H_2$ $cm^{-2}$ $s^{-1}$, and a mass-based TOF of 2.9 mmol $H_2$ $g^{-1}$ $s^{-1}$ (Fig. 4b). This enhancement arises from the coexistence of metallic Mo and $MoS_2$ domains, as well as sulfur vacancies that activate normally inert basal planes. In contrast, sulfur oversupply (M8.0–M9.0) induces structural disorder and reduces catalytic performance.  A number of prior DFT studies strongly support the structure-property relationships observed in this work. Kibsgaard *et al.* demonstrated that Mo-terminated edges are the dominant HER-active sites in $MoS_2$, while the basal plane remains largely inert, establishing that a reduction in edge density unavoidably weakens catalytic activity.[4] Similarly, Li *et al.* confirmed through combined DFT and experimental analysis that HER performance decreases sharply as edge exposure diminishes.[28] Komsa and Krasheninnikov further showed that moderate sulfur vacancies introduce localized Mo d-states that enhance electronic conductivity by creating metallic channels within the band structure,[29] while Ye *et al.* demonstrated that such vacancies can tune the hydrogen adsorption energetics toward thermoneutral values, yielding an optimal defect concentration for maximum HER activity.[52] Noh *et al.* also demonstrated that metallic Mo inclusions or partially sulfided $MoS_x$ domains create parallel conductive pathways that significantly lower charge-transport resistance in mixed-phase Mo/$MoS_2$ systems.[53] Moreover, Qiu *et al.* established that increased crystallinity improves carrier mobility but concurrently decreases the relative fraction of catalytically accessible edges,[54] while Eda and Maier showed that increased layer thickness strengthens interlayer coupling at the cost of reduced edge-to-area ratio, thereby diminishing HER activity.[55] Collectively, these DFT insights demonstrate that maximal HER activity in $MoS_2$ emerges not from any single structural parameter but from a synergistic balance between structural order, defect density, and electronic conductivity-precisely the interplay observed in our MBE-grown $MoS_2$ films. The Tafel slopes extracted from the polarization curves vary systematically with growth conditions (Table 1), providing insight into the HER kinetics in alkaline media. In alkaline

HER, the initial Volmer step involves water dissociation ($H_2O + e^- \rightarrow H^* + OH^-$) and is often kinetically limiting on $MoS_2$-based catalysts, making the availability of defect- and edge-rich sites as well as efficient electron transport critical. For the annealing series, the Tafel slope increases from 136 mV dec$^{-1}$ (MoS-T600) to 257 and 297 mV dec$^{-1}$ for MoS-T700 and MoS-T800, indicating slower apparent kinetics with increased crystallinity and stacking order. In contrast, the deposition-cycle series shows a minimum Tafel slope of 80 mV dec$^{-1}$ for MoS-N10, coinciding with the lowest charge-transfer resistance and an optimal balance between edge accessibility and conductivity. A similar trend is observed in the sulfur-supply series, where moderate sulfur deficiency (*e.g.*, MoS-M6.0) yields reduced Tafel slopes whereas sulfur-rich films exhibit larger values. To obtain a more physically meaningful descriptor of electron-transfer kinetics under operating conditions, we rely primarily on electrochemical impedance spectroscopy performed at HER-relevant potentials. The charge-transfer resistance ($R_{ct}$) extracted from EIS directly reflects the interfacial electron-transfer barrier during HER and varies systematically with sulfur stoichiometry, annealing temperature, and deposition cycles, whereas the solution resistance ($R_s$) remains constant. These $R_{ct}$ trends correlate consistently with Tafel slopes, ECSA, and spectroscopic indicators of defect density and stacking order, providing a robust basis for interpreting the role of electronic transport in governing HER activity.

The Nyquist plots were fitted using an equivalent circuit comprising Rs, $R_{ct}$, and a constant phase element. Owing to identical electrolyte and cell configurations, $R_s$ remains nearly constant across all samples (Figs. S5, S10, and S19), indicating a negligible contribution from solution resistance. In contrast, $R_{ct}$ varies systematically with growth conditions (Table 1), with lower values for moderately sulfur-deficient or optimally deposited films (*e.g.*, MoS-N10 and MoS-M6.0) and higher values for highly annealed or sulfur-rich samples. These trends, consistent with ECSA, Tafel slope, Raman edge signatures, and resistivity, indicate that the apparent HER kinetics arise from the combined effects of Volmer-step facilitation, interfacial charge transfer governed by effective conductivity, and stacking-induced modulation of edge accessibility.

Across all growth conditions, MBE-grown $MoS_2$ films exhibit a clear and consistent structure–activity correlation. By systematically varying sulfur stoichiometry, thickness, and annealing-driven structural evolution, MBE enables controlled tuning of crystallinity, defect density, and electronic transport, which can then be correlated with HER activity. Excessive crystallinity induced by high-temperature annealing or large deposition cycles reduces edge-site density and effective conductivity, whereas moderate structural disorder, introduced by controlled sulfur deficiency or intermediate cycle numbers, creates additional catalytic pathways through defect-derived active sites and mixed Mo/$MoS_2$ conductive domains. Consequently, optimal HER performance is achieved not under fully stoichiometric or highly crystalline conditions, but in an intermediate regime of moderate sulfur deficiency, where reduced charge-transfer resistance, favorable Tafel kinetics, and enhanced edge accessibility are simultaneously realized, an outcome enabled by the sub-monolayer flux precision and wafer-scale uniformity unique to MBE. In this regime, subtle interlayer sliding between $MoS_2$ sheets may further promote local charge redistribution and polarization, synergistically enhancing carrier mobility and catalytic turnover. Benchmarking against representative literature catalysts (Table S1) confirms that the best MBE-grown sample (MoS-N10) delivers competitive mass-normalized HER activity at ultralow loading under alkaline conditions. Notably, the successful epitaxial growth of atomically uniform $MoS_2$ films directly on Si substrates, rarely achieved in prior studies, opens new opportunities for integrating transition-metal dichalcogenide catalysts with semiconductor platforms, where the Si substrate can additionally facilitate interfacial charge transfer and further enhance electrocatalytic efficiency.

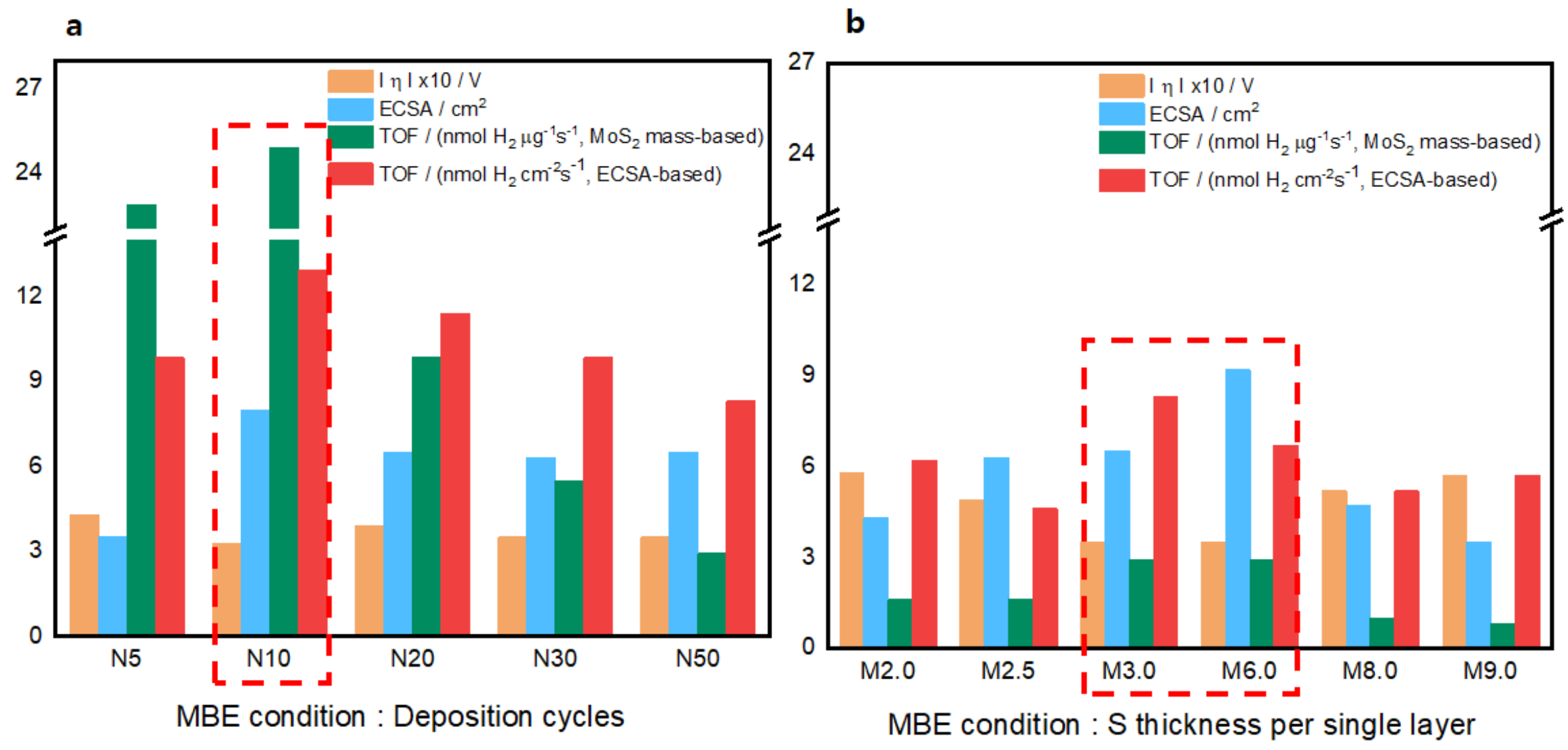

Fig. **4. a, b** Correlation of $MoS_2$ design parameters with HER performance.
**a** Deposition cycle effect on HER activity (N series). MoS-N10 exhibits the optimal balance of structural order, conductivity, and accessible surface area, delivering the lowest overpotential (-0.33 V at -10 mA $cm^{-2}$), the largest ECSA (8.0 $cm^2$), and the highest ECSA-based TOF of 13.0 nmol $H_2$ $cm^{-2}$ $s^{-1}$. Both lower (N5) and higher cycle numbers (N20–N50) result in inferior activity due to insufficient surface sites or excessive thickness-induced resistivity. **b** Sulfur stoichiometry effect on HER activity (M series). MoS-M6.0 achieves the best performance, with an overpotential of -0.35 V, an ECSA of 9.2 $cm^2$, and an ECSA-based TOF of 6.7 nmol $H_2$ $cm^{-2}$ $s^{-1}$, enabled by the coexistence of Mo and $MoS_2$ phases and sulfur vacancies that activate basal planes. In contrast, sulfur-rich films (M8.0, M9.0) show reduced activity, reflecting stacking disorder and diminished conductivity.

## Conclusions

We have demonstrated that the catalytic performance of MBE-grown $MoS_2$ can be tailored through controlled modulation of annealing temperature, deposition cycle number, and sulfur stoichiometry. High-temperature annealing (800 °C) and excessive deposition cycles (N50) enhance crystallinity but simultaneously reduce the electrochemically active surface area (ECSA, 3.5 $cm^2$) and HER activity ($\eta = -0.58$ V at −10 mA $cm^{-2}$), reflecting the loss of edge sites and decreased conductivity. In contrast, intermediate deposition cycles (N10) and moderate sulfur deficiency (M6.0) yield mixed-phase Mo + $MoS_2$ heterostructures featuring sulfur vacancies that activate the basal plane. These optimized films achieve overpotentials of −0.33 to −0.35 V, enlarged ECSA up to 9.2 $cm^2$, and mass-based TOF values exceeding 23 nmol $H_2$ $\mu g^{-1}$ $s^{-1}$, more than twice those of stoichiometric counterparts. Our results reveal that maximal HER performance does not emerge from fully crystalline $MoS_2$, but from a synergistic balance of structural order, defect density, and electrical conductivity. The coexistence of metallic Mo and semiconducting $MoS_2$ domains provides concurrent pathways for charge transport and catalytic turnover, establishing defect-controlled stoichiometry and growth kinetics as powerful levers for atomic-scale optimization. This correlation between sulfur stoichiometry, basal-plane activation, and electrochemical metrics clarifies how non-stoichiometric growth enhances intrinsic activity beyond that of ideal crystalline films. More broadly, this study defines a generalizable design principle for layered catalysts, where deliberate off-stoichiometry, layer thickness and controlled disorder yield optimal performance through the interplay of conductivity, polarization, and defect-mediated reactivity. Such an approach can be extended to other transition-metal dichalcogenides and van der Waals heterostructures, paving the way for next-generation electrocatalysts and energy-conversion materials engineered with atomic precision. The successful epitaxial growth of atomically uniform $MoS_2$ films on Si substrates marks a rare and significant achievement, enabling direct integration of transition-metal dichalcogenide catalysts with semiconductor platforms.

## Methods

*Growth of $MoS_2$ film*

The Mo-S samples were synthesized using the modulated elemental reactants method (MER) in a OCTOPLUS 350 molecular beam epitaxy (MBE) system (Dr. Eberl MBE-Komponenten GmbH).[51] The system comprises a load-lock chamber for wafer loading and unloading, as well as baking at low temperatures (with 8 wafer positions); a heated station chamber for substrate baking at higher temperatures, and a growth chamber for the actual sample synthesis. Silicon (Si) substrates with a native oxide layer (Dummy CZ-Si wafer 2-inch, thickness 500 ± 50 µm, 1-sided polished, p-type boron doped, MicroChemicals) were loaded without pretreatment and baked in the load-lock chamber at 150°C for 8 h to remove water and other contaminants (pressure in load-lock chamber better than $5x10^{-9}$ mbar).

Molybdenum pellets (Mo 99.95% pure ¼” diameter x ¼” long, Kurt J. Lesker company) were loaded into the vertical electron beam evaporator (EBVV, Dr. Eberl MBE-Komponenten GmbH) without a crucible liner. The EBVV is operated in the MBE chamber with a base pressure of ~$1x10^{-10}$ mbar, and a 40 volume % propane-1,2-diol / 60 volume % distilled water mixture of -15C is pumped through the liquid nitrogen shield by a chiller (CS 50W-FF-SA-15B-65, Van der Heijden Labortechnik GmbH) to cool the whole MBE growth chamber. All unused effusion cells have a stand-by temperature just below the evaporation starts to significantly reduce sulfur poisoning of the present elements.

Tin disulfide ($SnS_2$, 99.999%, Heeger Materials Inc., irregular pieces) was loaded into a pyrolytic boron nitride (pBN) crucible into a DECO effusion cell (Dr. Eberl MBE-Komponenten GmbH) with a Ga trapping cap unit (Dr. Eberl MBE-Komponenten GmbH) to catch possible SnS flux. Serving as a source of S, the $SnS_2$ effusion cell was heated to 390°C, undergoing thermal decomposition based on the following equation:

$$SnS_2(s) \rightarrow SnS(s) + S(g)$$

The less volatile SnS remained in the crucible, while the more volatile S was evaporated to the sample, thereby increasing the background pressure in the growth chamber to the low $10^{-9}$ mbar range. Note that a significantly smaller cell temperature was used compared to Shimada *et al.*,[56] and using a quadrupole mass spectrometer (QMG 250 M2 PRISMAPRO, Pfeiffer Vacuum Technology AG), only S (32 amu), $S_2$ (64 amu), and $S_3$ (96 amu) were detected, and no signal of longer S-chains could be detected.

The growth rates of Mo and S were calibrated using a quartz crystal microbalance (QCM) mounted in the MBE chamber. The QCM determines the growth rate by measuring the shift in the resonant frequency of the quartz crystal oscillator as a material's atoms condense on its surface. This rate is expressed in angstroms per second (Å/s), assuming a known material density, Z-factor, and sticking coefficient. Initially, the Mo shutter was opened, and all the other shutters closed to measure the Mo flux, which was typically around 0.05 Å/s or 0.6 x $10^{-2}$ atoms / $Å^2$ / s. The Mo rate was determined before the $SnS_2$ source was heated to 390°C, and after the $SnS_2$ source was heated to 390°C, and the pressure in the growth chamber increased due to S to verify for subsulfide formation.[57-59]

As the sticking coefficient of S is extremely low on a water-cooled QCM, the S flux was measured by co-deposition of Mo, such that the final rate also takes into account the sticking coefficient of S to Mo, similar to what is done for the calibration of atomic oxygen using silver-coated quartz crystal.[60, 61] Therefore, the $SnS_2$ (serving as an S source) shutter was opened concurrently with the Mo shutter, and the combined growth rate of Mo and S ($R_{Mo+S}$) was measured. The S shutter was then closed, and the Mo shutter was opened alone while maintaining the QCM sensor in the configuration used for S growth rate measurement, the standalone Mo growth rate $R_{Mo}$. The S growth rate was obtained by subtracting $R_{Mo}$ from $R_{Mo+S,}$ accounting for their differential mass accumulation due to S incorporation. The S rate was measured to be ~0.05 Å/s or 0.2 x $10^{-2}$ atoms / Å$^2$ / s. Care should be taken that there is a maximum amount of S that can stick to the co-evapoerated Mo, and too large S flux will not be detected. Similar, S can diffuse into previous deposited materials onto the QCM, which depending on the previously deposited materials can result in variations of the observed rate with larger S fluxes. Furthermore, the actual S flux depends also on the amount of $SnS_2$ source material loaded in the crucible and/or the time the source has been used, and a slightly higher source temperature might be needed. If the $SnS_2$ shutter is opened/closed, a clear increase/decrease of the chamber pressure can be observed.

Ultrathin Mo ($d_{Mo}$) and S ($d_S$) layers were deposited onto the natively oxidized Si substrates using the shuttering sequence of Mo and S molecular beams, where the deposition times of Mo and S were calculated from their unit cell parameters (1 monolayer $MoS_2$ needs 0.114 Mo atoms / Å$^2$ and 0.228 S atoms / Å$^2$). The sequence was repeated N times to grow an N-layer-thick $MoS_2$ film. The wafer was rotated during the growth and annealing, azimuthally at 15 rotations per minute (rpm) to ensure uniform deposition and temperature distribution across the surface. After growth, the shutters of Mo and S were both closed, and both cells were cooled down, such that no S and Mo flux was present. Subsequently, the deposited Mo/S layers were heated in the growth chamber at 10°C/min to the annealing temperature, where they were kept for 1 hour. Finally, the sample was cooled to room temperature at a rate of 10°C/min.

*Material characterizations*

Using reflection high-energy electron diffraction (RHEED, RHEED-15, Staib Instruments GmbH) the samples were characterized during the deposition and annealing phase of the growth. A collimated beam of electrons, accelerated to 15 keV with a filament current of 1.5

A, was directed onto the sample surface in a grazing-incidence reflection geometry. The resulting diffraction pattern, representing the reciprocal lattice, was projected onto a phosphor screen and recorded using an external digital camera. Prior to material deposition, the RHEED images of the silicon wafers (lattice parameter: 5.43 Å) were acquired for calibration. The acquired RHEED images were subsequently analyzed using PyRHEED, an open-source Python-based software package designed for the analysis and simulation of RHEED data. PyRHEED was employed to process these images, enabling the extraction of structural information from the diffraction patterns, including lattice constants.[62]

After taking the 2-inch wafer out of the chamber, the wafer was cut in small pieces for the different experiments performed.

Scanning probe microscopy was conducted on an Asylum Research (Oxford Instruments, Goleta, CA, USA) Cypher S atomic force microscopy (AFM) instrument with an air temperature controller (Cypher ATC) using multi75-G probes (BudgetSensors).

Differential reflection spectroscopy of the grown thin films was measured using a home-built setup. The setup consists of a compact CCD spectrometer (Thorlabs, Inc. CCT10), which is coupled with a tungsten-halogen light source (Thorlabs, Inc. OSL2IR) into a fiber optic reflection probe (RP28, Thorlabs, Inc.), which is coupled to a pair of identical plano-convex lenses (LA4306-ML, Thorlabs, Inc.) to illuminate the sample or bare Si wafer and collect the back-reflected light.[63]

The crystal structures of Molybdenum sulfide were determined by X-ray diffraction and X-ray reflectivity using a Rigaku SmartLab SE diffractometer equipped with Cu Kα radiation ($\lambda$ = 1.5406 Å). Diffraction data were collected over a 2θ range of 10°–60° at a scan rate of 2° $min^{-1}$. Raman spectra of $MoS_2$ samples were performed using confocal microscope Raman spectroscopy (HEDA, WEVE) with a 532 nm excitation laser operated at 25 mW. The laser beam was focused onto the sample through a 20× objective lens. The resistivity was measured using a commercial four-point probe resistivity measurement system (CMT-SR2000N, AIT). Resistivity measurements were performed at room temperature. The reported values correspond to the combined resistivity of the $MoS_2$ and the substrate ($MoS_2$ + Si wafer), with the substrate alone having a resistivity of 14.6 Ω·cm. X-ray absorption spectra at Mo K-edge for references and sample films were acquired in the energy range of 19.80–21.00 keV using synchrotron radiation at beamline 8C of the Pohang Light Source (PLS), which provides a flux of $2 \times 10^{12}$ photons·$s^{-1}$ at 100 mA and 3 GeV. Extended X-ray absorption fine structure spectra at the Mo K-edge were measured at ambient temperature in fluorescence mode using a silicon drift detector (SDD). The $k^3$-weighted EXAFS spectra were Fourier transformed over the range 2.5–14.0 $Å^{-1}$. The curve fitting was performed with WinXAS 3.1. Cross-sectional $MoS_2$ lamellae were prepared using a focused ion beam–scanning electron microscope (Helios 5 UX, Thermo Fisher Scientific). To minimize ion-beam-induced damage during milling, a protective carbon capping layer was deposited prior to sectioning. The lamellae were transferred onto Cu TEM grids for analysis. The images of scanning transmission electron microscopy were obtained using a double Cs-corrected high-resolution transmission electron microscope (JEM-ARM200F, JEOL) operated at 200 kV, following pre-irradiation by the electron beam to stabilize the sample.

*Electrochemical characterizations*

All electrochemical measurements were performed in a glass cell containing purified 1.0 M KOH (SAMCHUN, 95.0 %) solution treated with $Ni(OH)_2$ using a potentiostat (ZIVE MP2, WonA Tech).[64] The measurements were conducted in a three-electrode configuration with a graphite rod as the counter electrode, a Hg/HgO (1.0 M NaOH) electrode as the reference electrode, and $MoS_2$ films as the working electrode. The working electrode was prepared by cutting the wafer samples into 1 × 1 $cm^2$ pieces. A protective layer was applied by spin-coating a 0.3 mg $cm^{-2}$ Nafion (Fisher Scientific, D-521 5 wt% dispersion) onto the surface of $MoS_2$ at 1000 rpm for 60 s.[65] The coated film was dried in air for 30 min, followed by vacuum oven drying at 50 ℃ for 30 min. Cyclic voltammetry was conducted at a scan rate of 5 mV $s^{-1}$ in a potential range of -0.9 V to 0 V (vs. RHE) at room temperature with stirring at 150 rpm. Electrochemical hydrogen evolution reaction (HER) measurements were conducted using a standard three-electrode configuration. LSV was recorded under identical experimental conditions for each sample. To evaluate measurement reproducibility and mitigate electrochemical noise, each representative sample was measured independently three times, using freshly prepared electrodes for each run. The independently acquired LSV curves were first corrected for iR drop and then averaged point-by-point to obtain a representative polarization curve. To suppress high-frequency noise intrinsic to LSV measurements, a smoothing procedure was applied to the averaged dataset, while preserving the overall polarization trend and kinetic features. The residual standard deviation between the raw averaged LSV data and the corresponding smoothed trend was calculated and used as a quantitative indicator of measurement uncertainty. For the MoS-T800 sample, for example, the residual standard deviation was determined to be ±0.04 mA $cm^{-2}$, which reflects the noise level of the electrochemical measurement. This value is reported in the Supporting Information (Fig. S20) and serves as a statistical metric analogous to error bars for continuous LSV data. The CV data were corrected to compensate for iR drop caused by solution resistance through EIS measurement. The electrochemical surface area was compared by measuring the double-layer capacitance ($C_{dl}$) from current difference at 0.71 V (vs. RHE) in cyclic voltammetry scans performed at 20, 40, 60, and 80 mV $s^{-1}$ within the non-Faradaic potential range of 0.66 to 0.76 V (vs. RHE). Specific capacitance (Cs) of 40 μF $cm^{-2}$ was assumed for the calculation of ECSA.[66] The TOF was determined by dividing the current (converted to $H_2$ mol $s^{-1}$ *via* Faraday's law) at –0.40 V (vs. RHE) by the loading mass and ECSA-derived surface area, allowing for a quantitative evaluation of the intrinsic activity of the catalysts.

The table of contents graphic was made using Microsoft Office PowerPoint.

**Data availability**

Data underlying the figures and conclusions of this work are publicly available *via* Zenodo *via* https://doi.org/10.5281/zenodo.17567341

The used (but freshly grown) $MoS_2$ samples presented in this work are available as 'Open samples' upon reasonable request.

## Acknowledgements

EJ and YKL acknowledge financial support from Dankook University in 2025 and the National Research Foundation of Korea (NRF-2022R1A2C2093257). VMP, MHK, and TV acknowledge financial support of the European Union by the ERC-STG project 2D-sandwich (Grant No 101040057) and the assistance provided by the Operational Programme Johannes Amos Comenius of the Ministry of Education, Youth and Sport of the Czech Republic, within the frame of project Ferroic Multifunctionalities (FerrMion) [Project No. CZ.02.01.01/00/22_008/0004591], co-funded by the European Union.

## Supporting Information

The Supporting Information is available free of charge at xxx

- Supporting Information includes characterization details for all the samples including XRR, XRD, differential reflectance spectra, RHEED, AFM and TEM images, cyclic voltammetry curves, Tafel and Nyquist plots, the data processing workflow for HER polarization curves, and catalytic performance comparison.

Table 1 Structural parameters, growth conditions, and electrochemical performance of MBE-grown $MoS_2$ samples

| Sample ID | $MoS_2$ growth condition | | | Raman ratio $A_{1g}/E_{2g}$ | Resistivity / Ω·cm | ECSA / $cm^2$ | MoSx loadings / μg $cm^{-2}$ | η at -10 mA $cm^{-2}$ / $V_{RHE}$ | Tafel slope /mV $dec^{-1}$ | $R_{ct}$ / ohm $cm^2$ | ECSA-based TOF at -0.4 $V_{RHE}$ / nmol $H_2$ $cm^{-2}s^{-1}$ | MoSx mass-based TOF at -0.4 $V_{RHE}$ / nmol $H_2$ $μg^{-1}s^{-1}$ |
|---|---|---|---|---|---|---|---|---|---|---|---|---|
| | Annealing Temp. / °C | Number of deposition cycles | S-layer thickness / 0.1 nm S per 0.3 nm Mo | | | | | | | | | |
| MoS-T600 | 600 | | | 2.41 | 15.98 | 6.7 | 24.7 | -0.46 | 136 | 98.4 | 5.7 | 1.6 |
| MoS-T700 | 700 | 50 | 9.0 | 2.34 | 16.52 | 6.5 | 24.7 | -0.48 | 257 | 113.0 | 5.2 | 1.4 |
| MoS-T800 | 800 | | | 2.29 | 19.26 | 3.5 | 24.7 | -0.58 | 297 | 193.3 | 5.7 | 0.8 |
| MoS-N5 | | 5 | | 1.01 | 7.75 | 4.5 | 1.9 | -0.43 | 161 | 136.5 | 9.9 | 22.9 |
| MoS-N10 | | 10 | | 1.63 | 8.99 | 8.0 | 3.7 | -0.33 | 80 | 52.8 | 13.0 | 24.9 |
| MoS-N20 | 800 | 20 | 3.0 | 1.85 | 11.08 | 6.5 | 7.4 | -0.39 | 105 | 76.9 | 11.4 | 9.9 |
| MoS-N30 | | 30 | | 1.78 | 11.40 | 6.3 | 11.1 | -0.35 | 93 | 59.0 | 9.9 | 5.5 |
| MoS-N50 | | 50 | | 1.99 | 12.45 | 6.5 | 18.5 | -0.35 | 114 | 64.0 | 8.3 | 2.9 |
| MoS-M2.0 | | | 2.0 | 1.70 | 9.01 | 4.3 | 17.5 | -0.58 | 484 | 161.2 | 6.2 | 1.6 |
| MoS-M2.5 | | | 2.5 | 1.97 | 9.50 | 6.3 | 18.0 | -0.49 | 253 | 104.5 | 4.6 | 1.6 |
| MoS-M3.0 | 800 | 50 | 3.0 | 1.99 | 12.45 | 6.5 | 18.5 | -0.35 | 114 | 64.0 | 8.3 | 2.9 |
| MoS-M6.0 | | | 6.0 | 2.05 | 15.09 | 9.2 | 21.6 | -0.35 | 91 | 45.5 | 6.7 | 2.9 |
| MoS-M8.0 | | | 8.0 | 2.24 | 17.14 | 4.7 | 23.7 | -0.52 | 223 | 124.5 | 5.1 | 1.0 |
| MoS-M9.0 | | | 9.0 | 2.29 | 19.26 | 3.5 | 24.7 | -0.58 | 297 | 193.3 | 5.7 | 0.8 |

# Supplementary Information

## Controlling Mixed Mo/$MoS_2$ Domains on Si by Molecular Beam Epitaxy for the Hydrogen Evolution Reaction

Eunseo Jeon[1,*], Vincent Masika Peheliwa[2,3,*], Marie Hrůzová Kratochvílová[2], Tim Verhagen[2,†], Yong-Kul Lee[1,†]

[1]Laboratory of Advanced Catalysis for Energy and Environment, Department of Chemical Engineering, Dankook University, Yongin 16890, South Korea

[2]Institute of Physics of the Czech Academy of Sciences, Prague 182 00, Czech Republic

[3]Faculty of Mathematics and Physics, Charles University, Prague 121 16, Czech Republic

*Equally contributed, †To whom all correspondence should be addressed: verhagen@fzu.cz; yolee@dankook.ac.kr

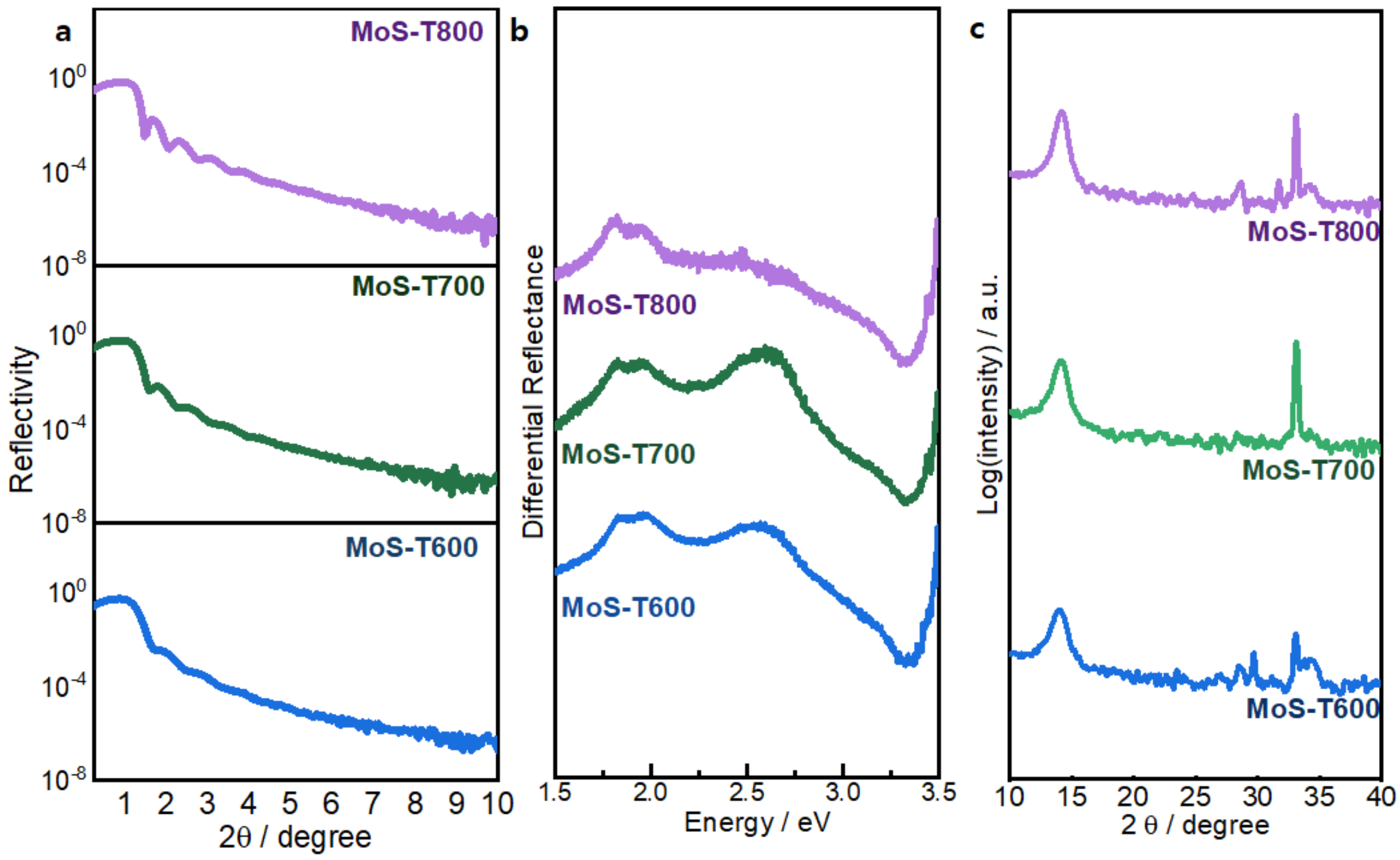

Figure S1. **a, b, c** Structural evolution of MBE-grown $MoS_2$: XRR patterns, differential reflectance spectra, and log-scaled XRD pattern of $MoS_2$ thin films grown with different annealing temperature.

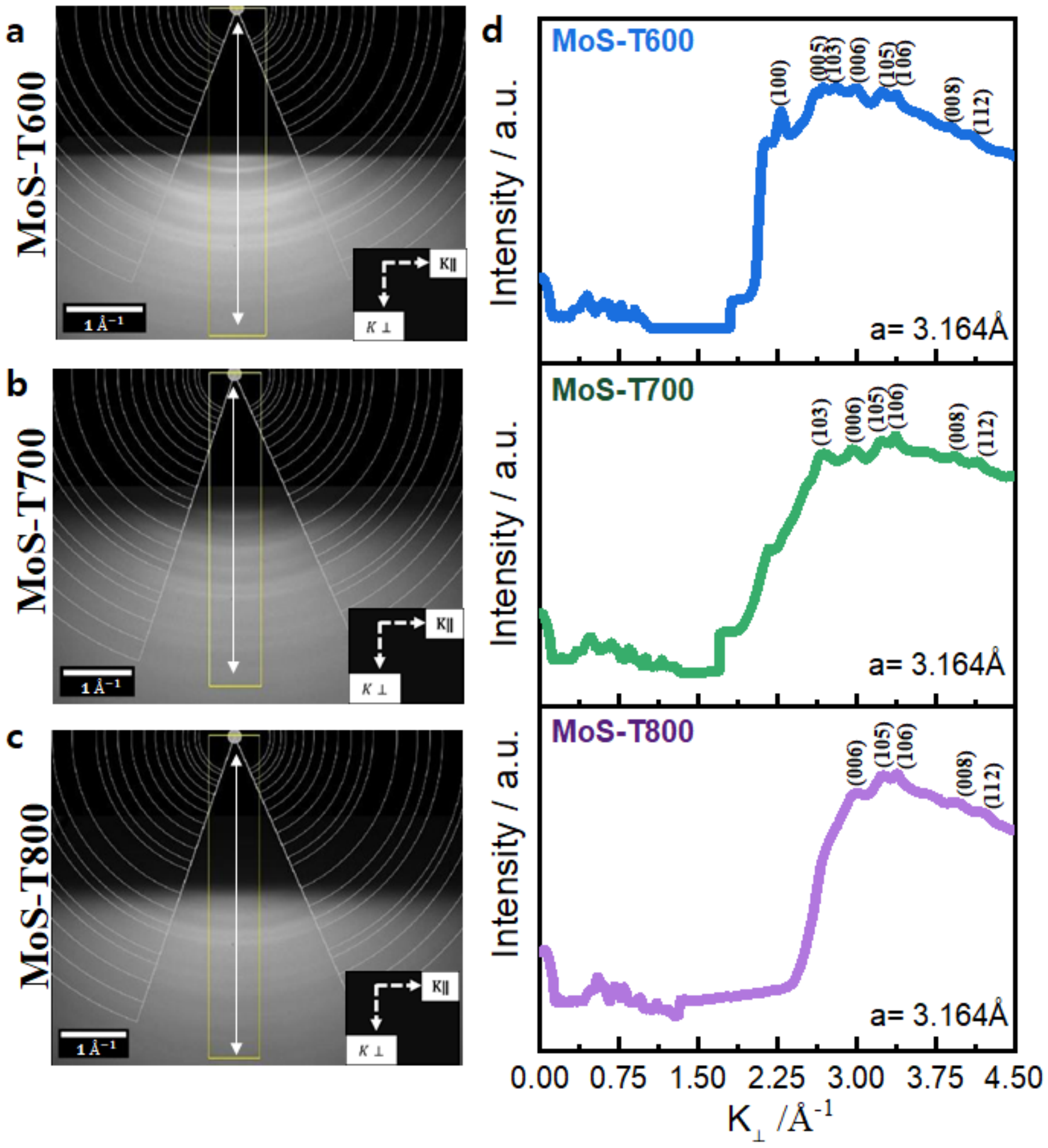

Figure S2. **a-d** Structural and surface characterization of $MoS_2$ thin films. RHEED patterns for MoS-T600 (**a**), T700 (**b**), and 800 (**c**) after annealing at 20 °C. **d** Intensity profiles as a function of momentum transfer perpendicular to the substrate $K_{\perp}$ along the white line shown in (**a-c**). The visible peaks are labeled according to the 2H- $MoS_2$ structure and the lattice parameter *a* is determined.

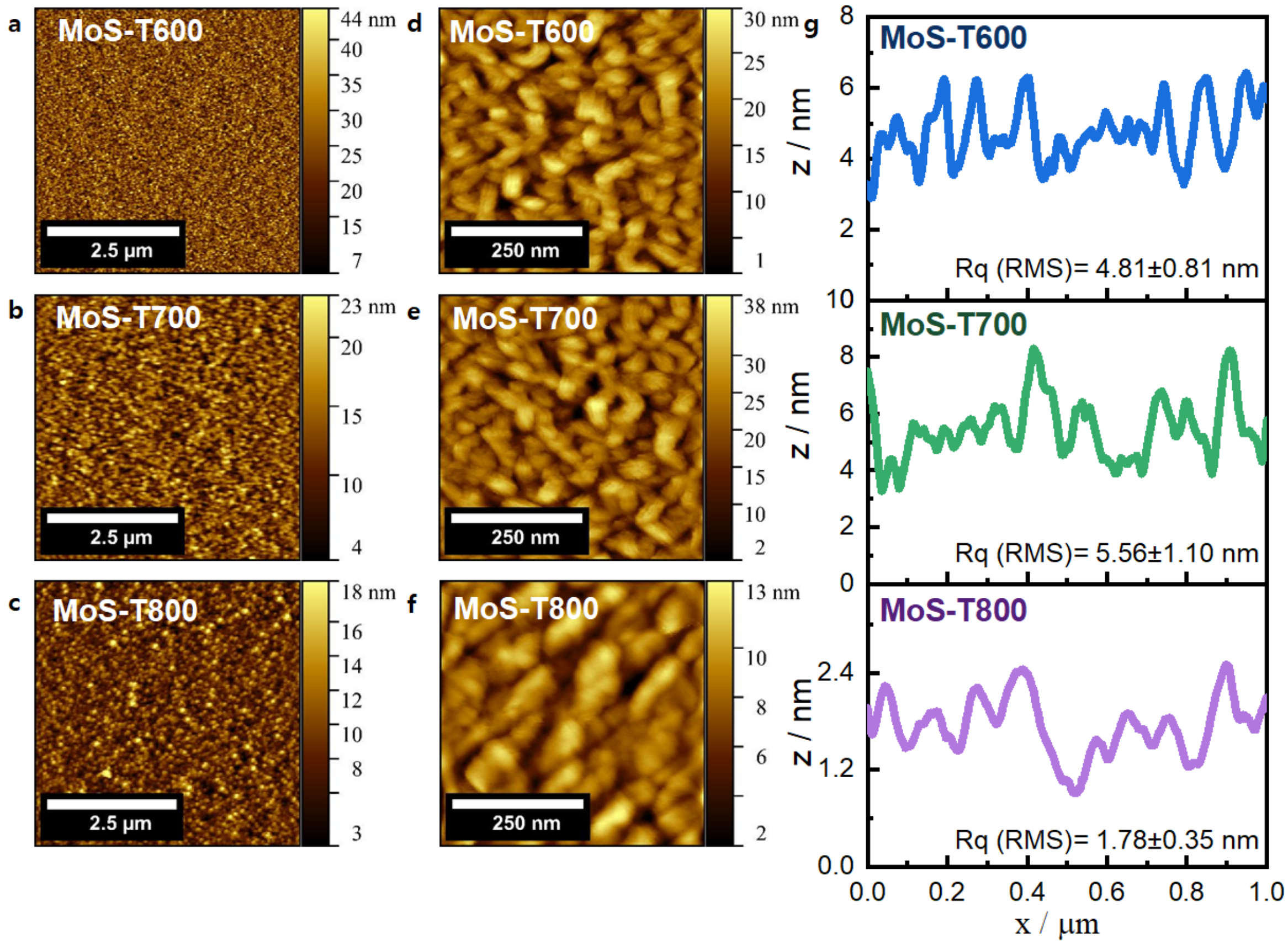

Figure S3. **a-f** AFM topography images showing the large scale (**a-c**) and small scale (**d-f**) surface morphology. **g** Surface roughness comparison of MoS-T600, MoS-T700, and MoS-T800. The height profile and shown surface root mean square roughness *Rq (RMS)* were obtained from the full images (**d-f**).

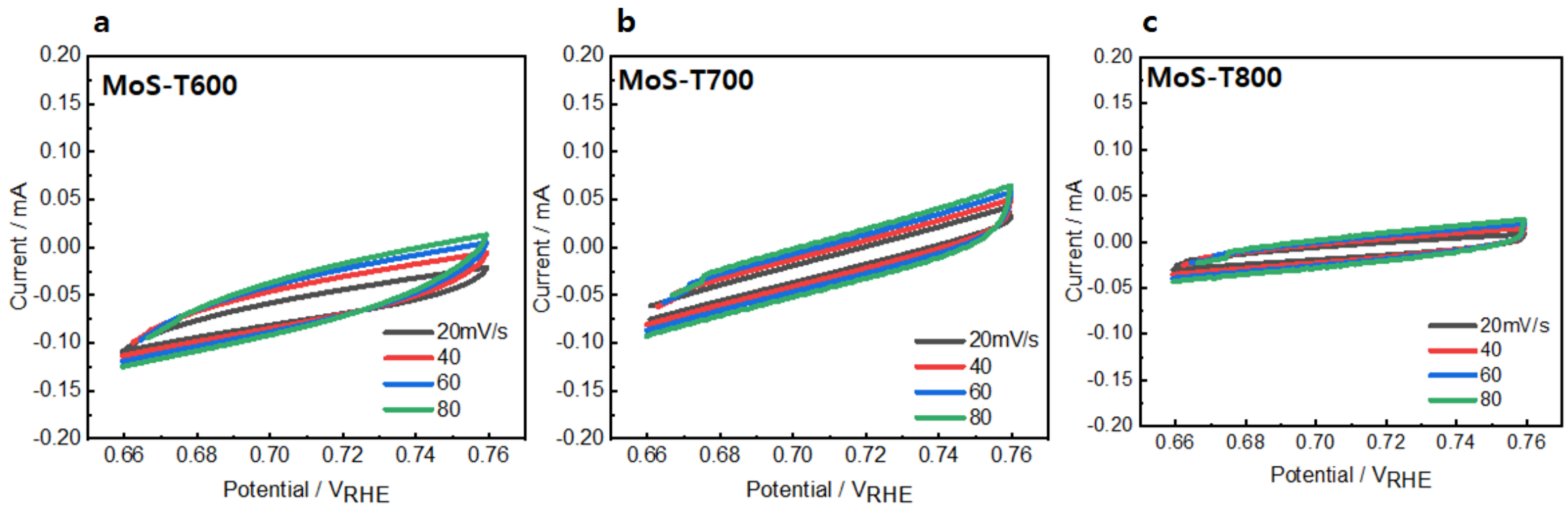

Figure S4. **a-c** Cyclic voltammetry curves of MoS-T600 (**a**), T700 (**b**), and T800 (**c**) measured at various scan rates (20, 40, 60, 80 mV $s^{-1}$) from 0.66 V to 0.76 V (vs. RHE).

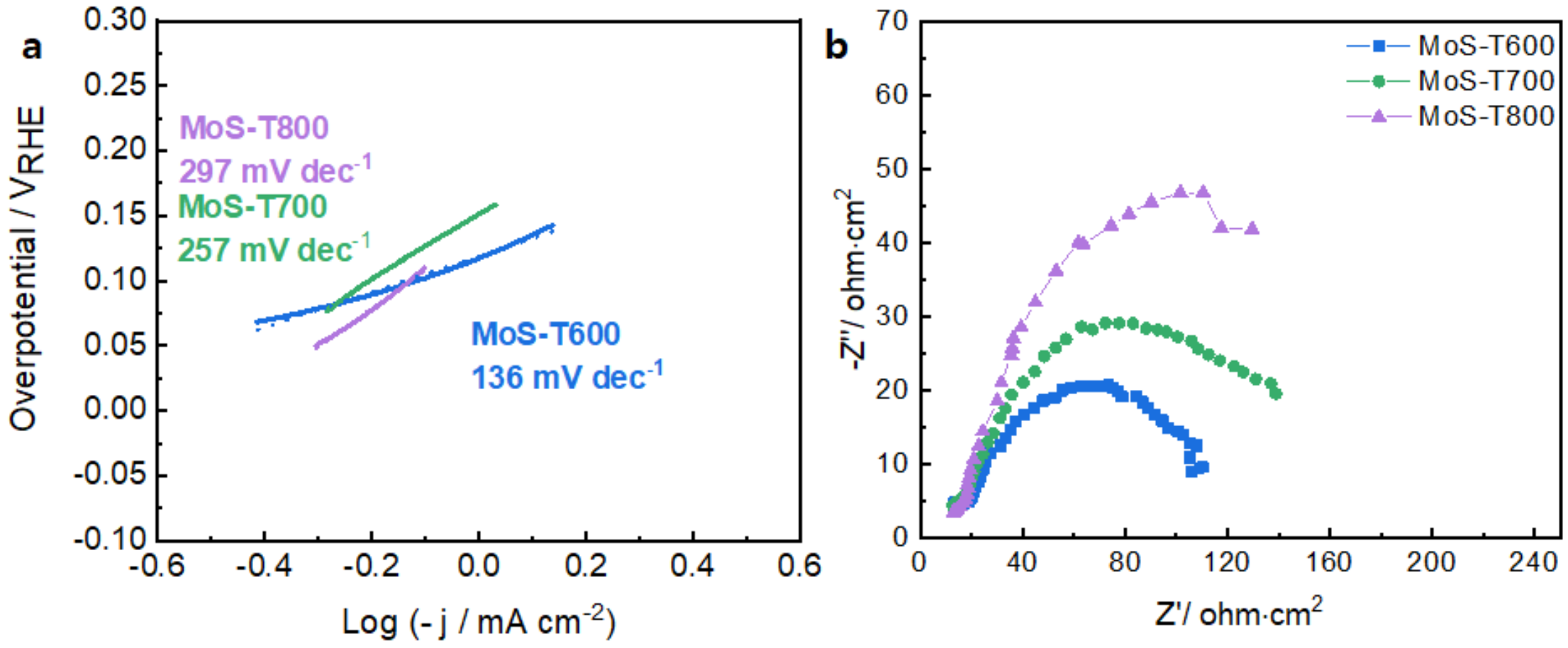

Figure S5. **a, b** Tafel plot and Nyquist plot for $MoS_2$ electrodes with different annealing temperatures. The EIS measurements were performed at –0.2 V (vs. RHE) using a frequency range of 100 kHz to 0.1 Hz.

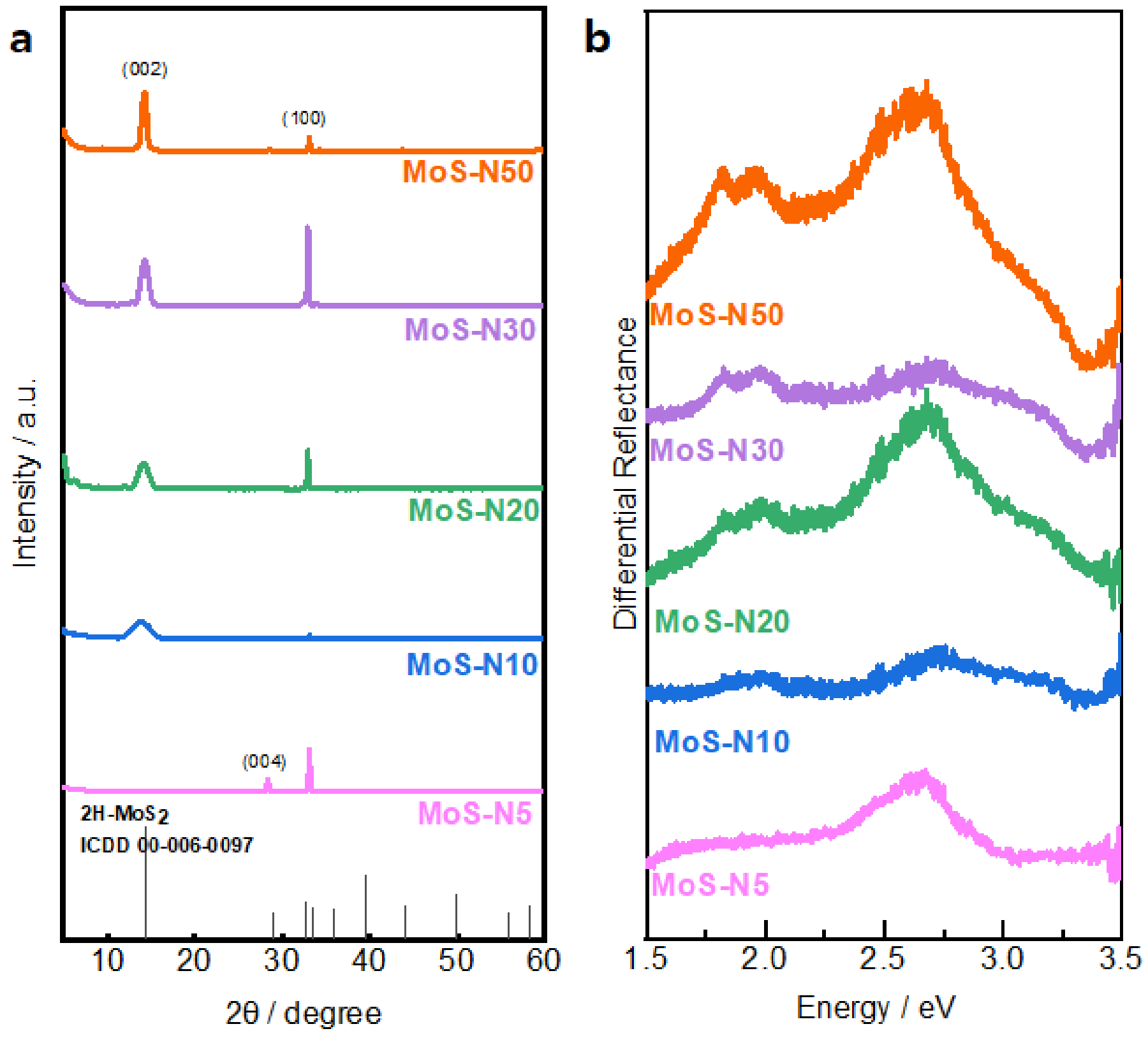

Figure S6. **a, b** Structural evolution of MBE-grown $MoS_2$: XRD patterns and differential reflectance spectra of $MoS_2$ thin films deposited with 5, 10, 20, 30, and 50 cycles.

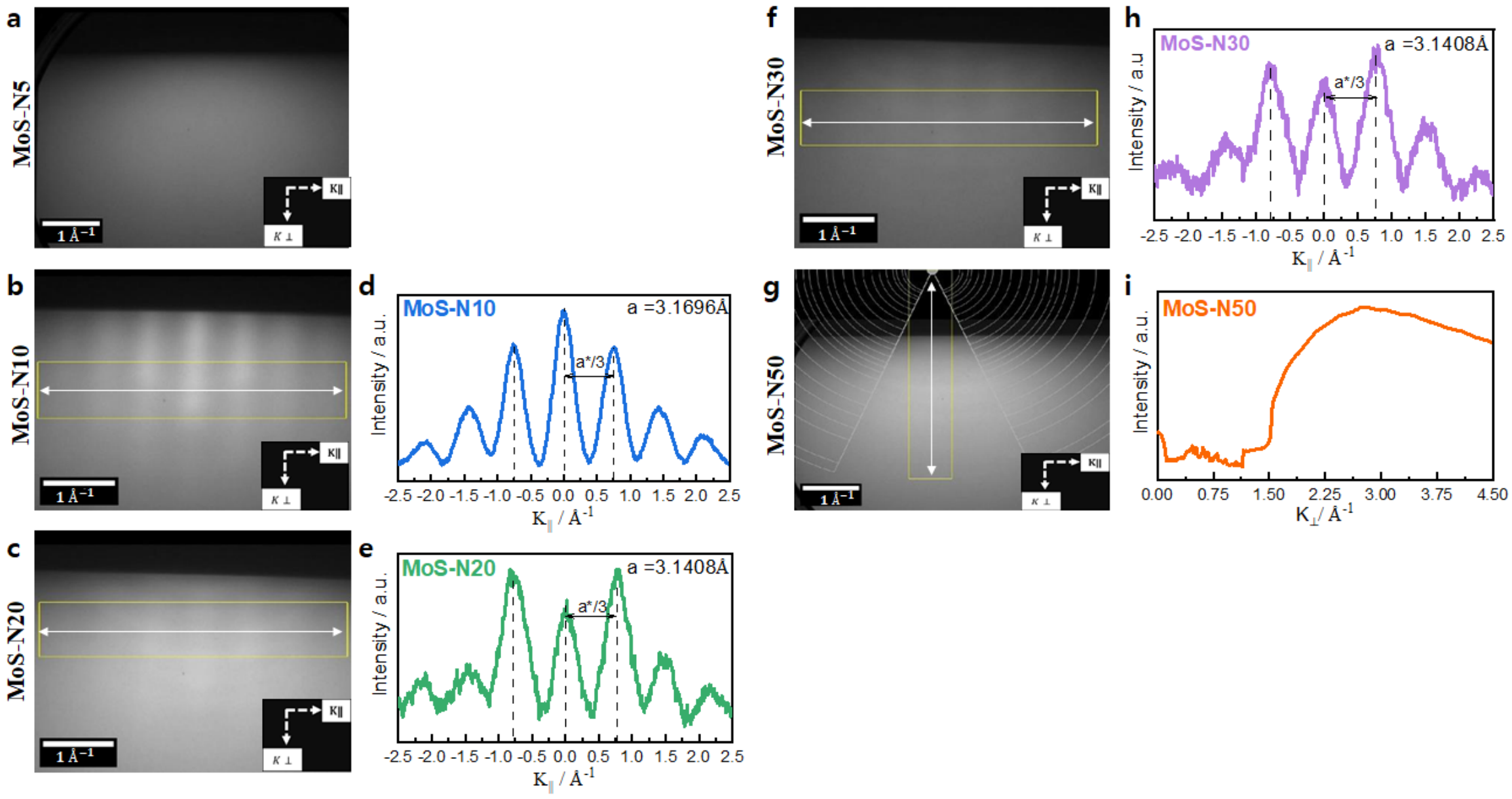

Figure S7. **a, b, c, f, g** RHEED patterns for the MoS-N5(**a**), N10 (**b**), N20 (**c**), N30 (**f**), and N50 (**g**). MoS-N5 shows no features, MoS-N10, MoS-N20, MoS-N30 show streaks, and MoS-N50 shows RHEED rings. **d, e, h** Intensity profiles as a function of momentum transfer parallel to the substrate $K_{\parallel}$ along the white line shown in panels (**b, c, f**). **i** Intensity profiles as a function of momentum transfer perpendicular to the substrate $K_{\perp}$ along the white line shown in panel (**g**).

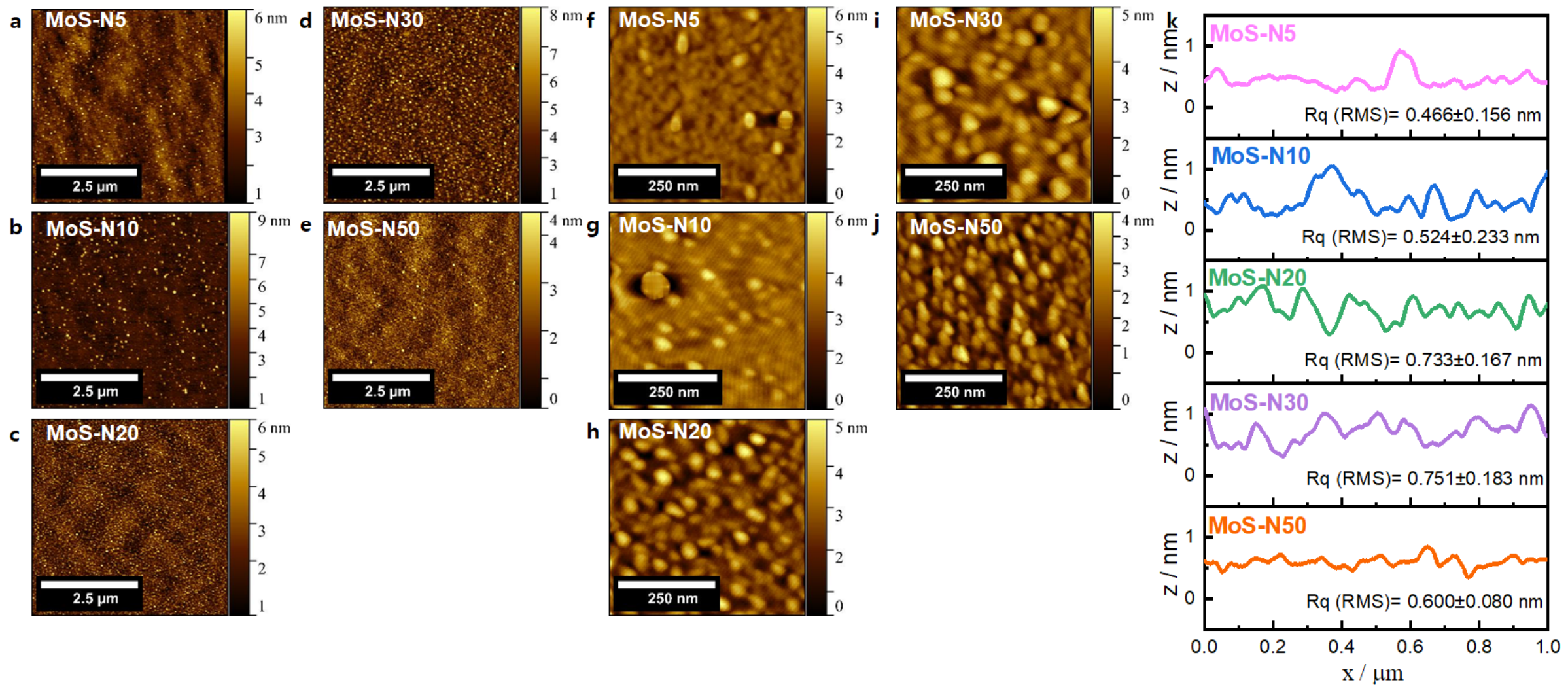

Figure S8. **a-j** AFM topography images showing the large scale (**a-e**) and small scale (**f-j**) surface morphology. **k** Surface roughness comparison for MoS-N5, MoS-N10, MoS-N20, MoS-N30, and MoS-N50. The height profile and shown surface root mean square roughness Rq (RMS) were obtained from the full images (**f - j**).

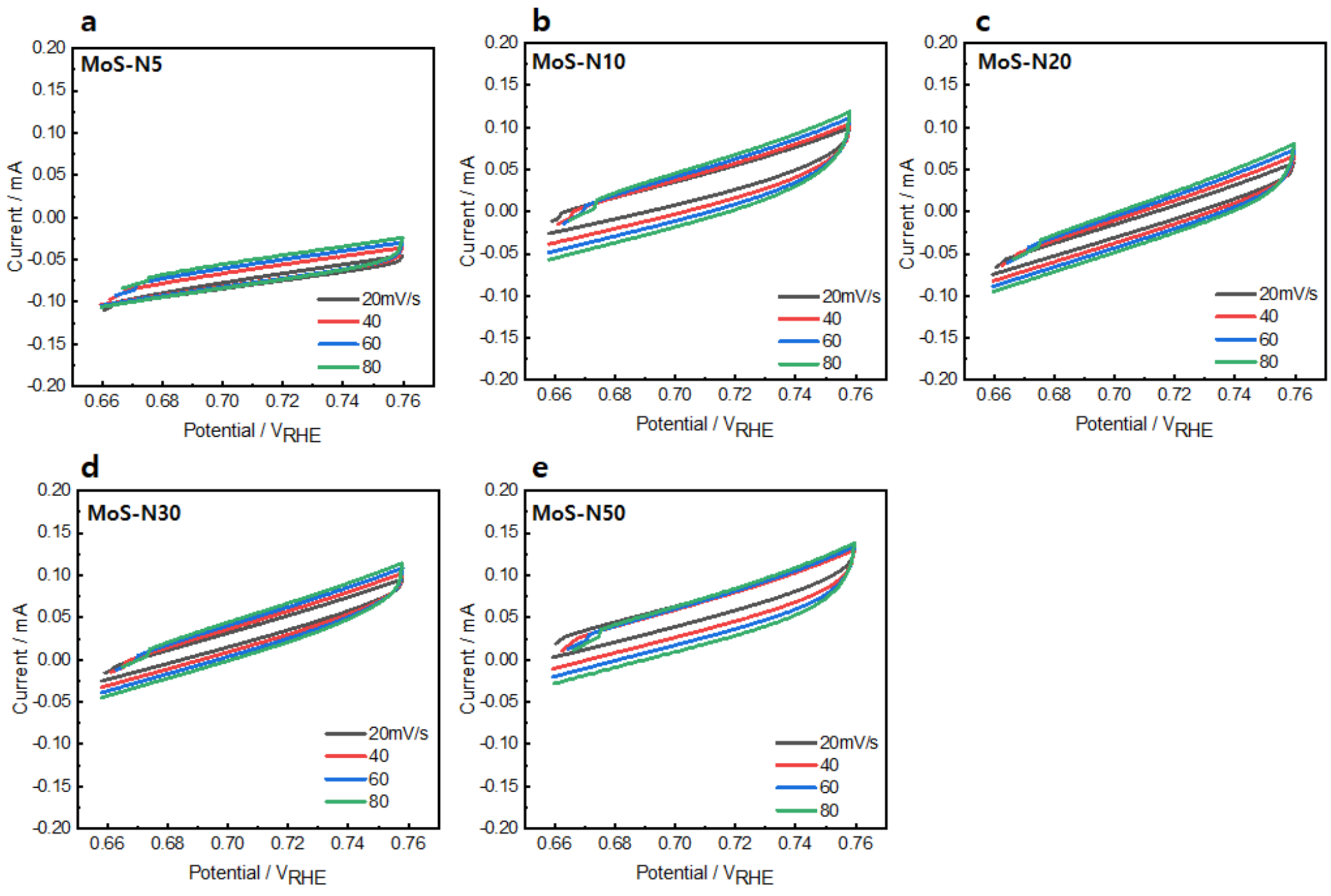

Figure S9, **a-e** Cyclic voltammetry curves of MoS-N5 (**a**), N10 (**b**), N20 (**c**), N30 (**d**), and N50 (**e**) measured at various scan rates (20, 40, 60, 80 mV $s^{-1}$) from 0.66 V to 0.76 V (vs. RHE).

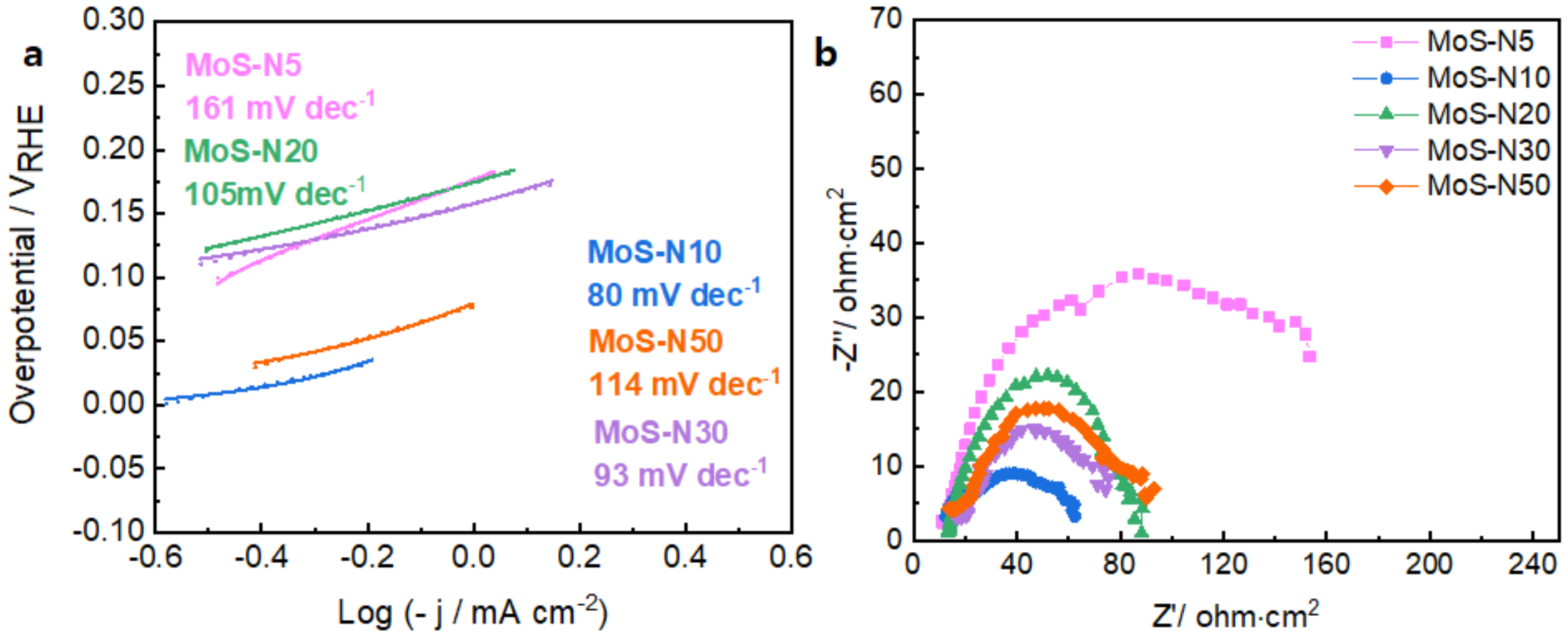

Figure S10. **a, b** Tafel plot and Nyquist plot for $MoS_2$ electrodes with different deposited cycles.

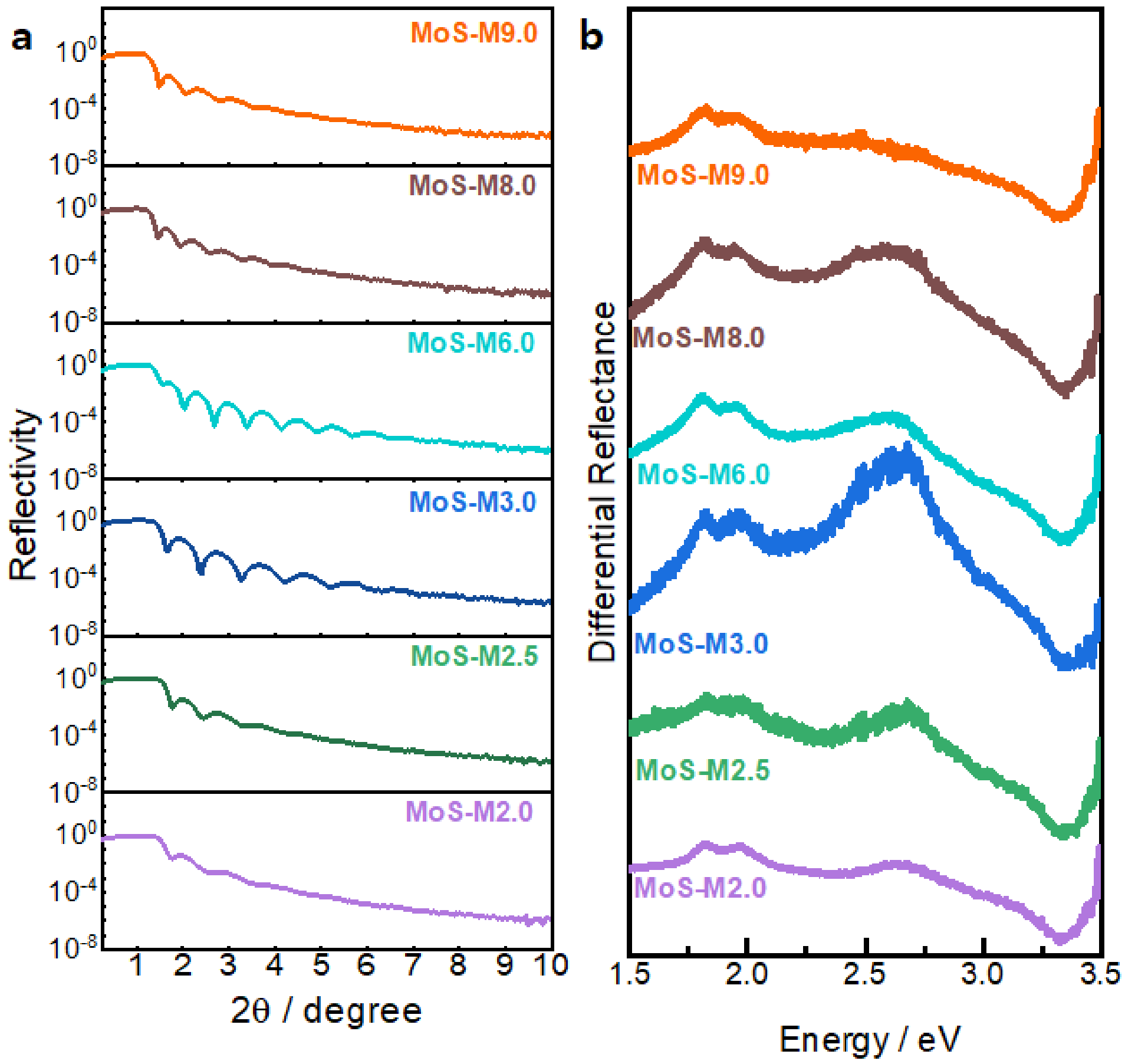

Figure S11. **a, b** Structural evolution of MBE-grown $MoS_2$: XRR patterns and differential reflectance spectra of $MoS_2$ thin films grown with varying sulfur thickness (2.0–9.0).

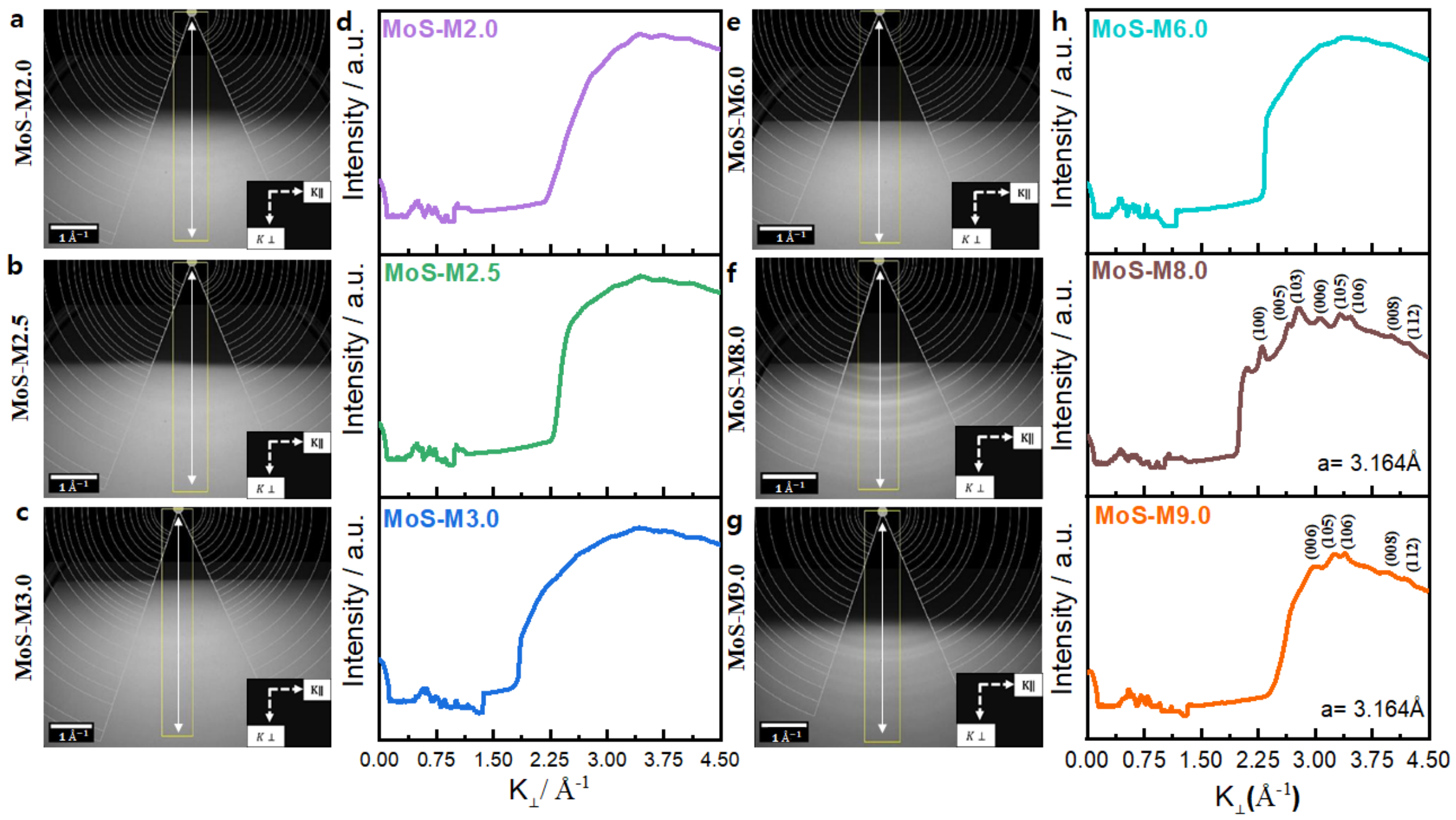

Figure S12. **a, b, c, e, f, g** RHEED patterns for MoS-M2.0 (**a**), MoS-M2.5 (**b**), MoS-M3.0 (**c**), MoS-M6.0 (**e**), MoS-M8.0 (**f**), and MoS-M9.0 (**g**). **d, h** Intensity profiles as a function of momentum transfer perpendicular to the substrate $K_{\perp}$ along the white line shown in panels (**a, b, c, e, f** and **g**). The diffraction peaks for MoS-M8.0 (**f**), and MoS-M9.0 (**g**) are labeled according to the 2H- $MoS_2$ structure.

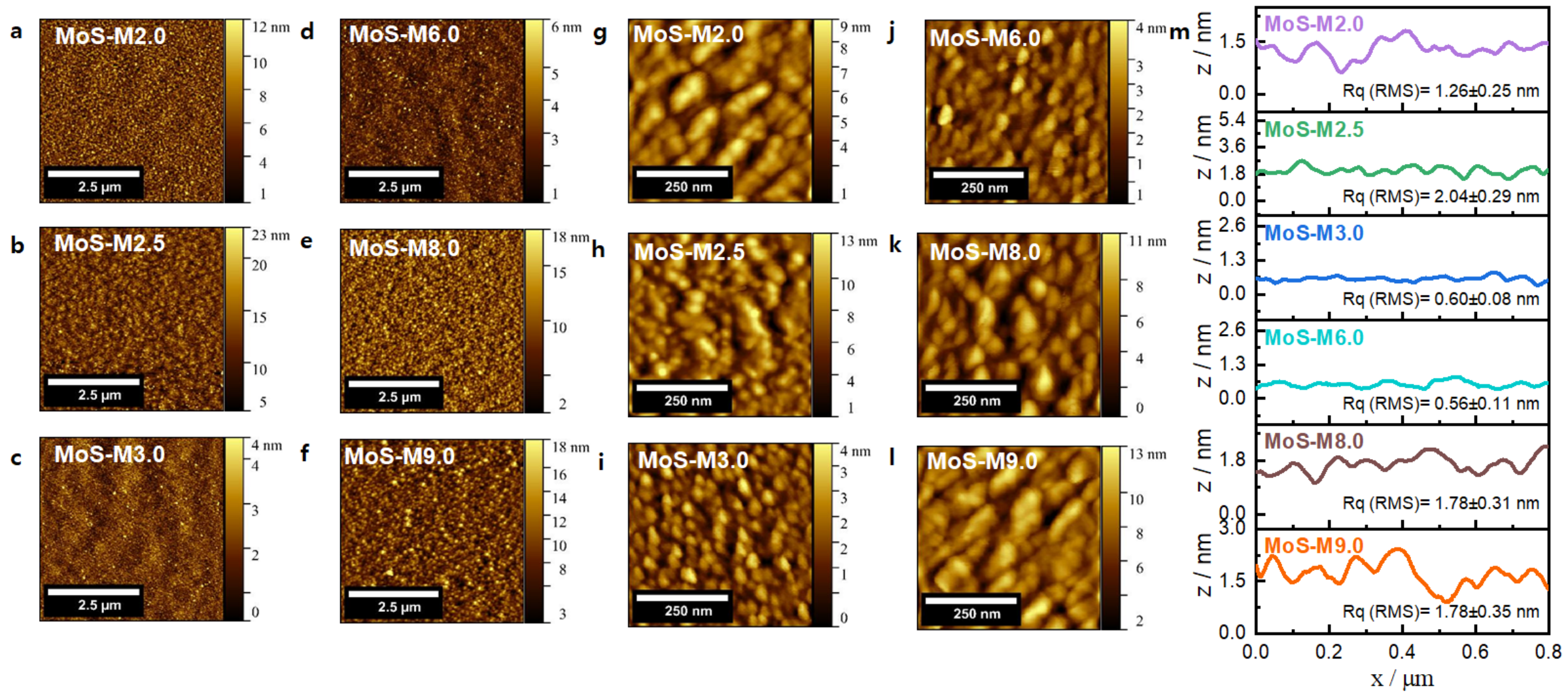

Figure S13. **a-l** AFM topography images showing the large scale and small scale surface morphology for MoS-M2.0 (**a, g**), MoS-M2.5 (**b, h**), MoS-M3.0 (**c, i**), MoS-M6.0 (**d, j**), MoS-M8.0 (**e, k**), and MoS-M9.0 (**i, l**). **m** The height profile and shown surface root mean square roughness *Rq (RMS)* were obtained from the full small scale images (**g-l**).

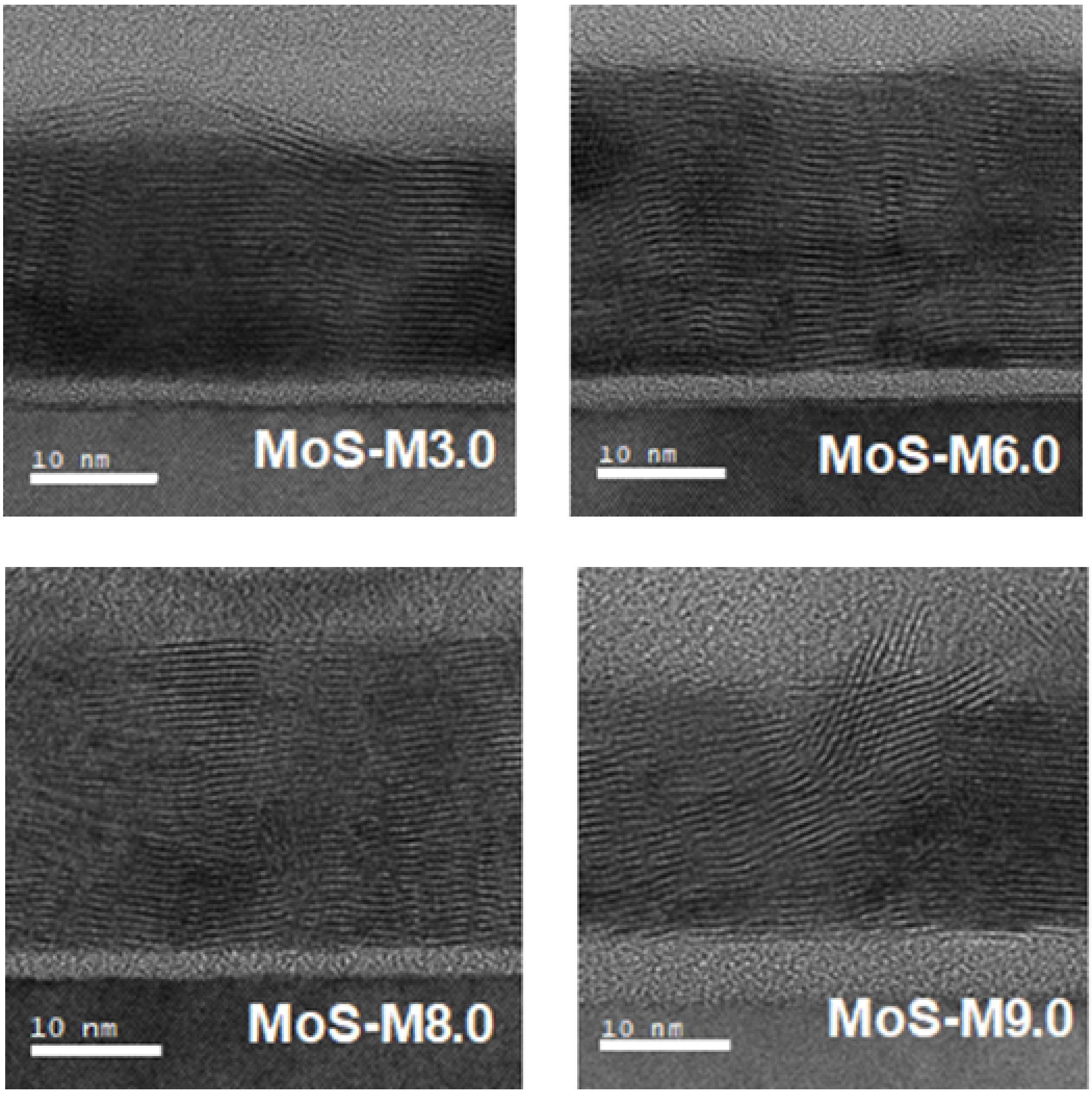

Figure S14. TEM images of $MoS_2$ films grown with varying sulfur thickness (3.0–9.0).

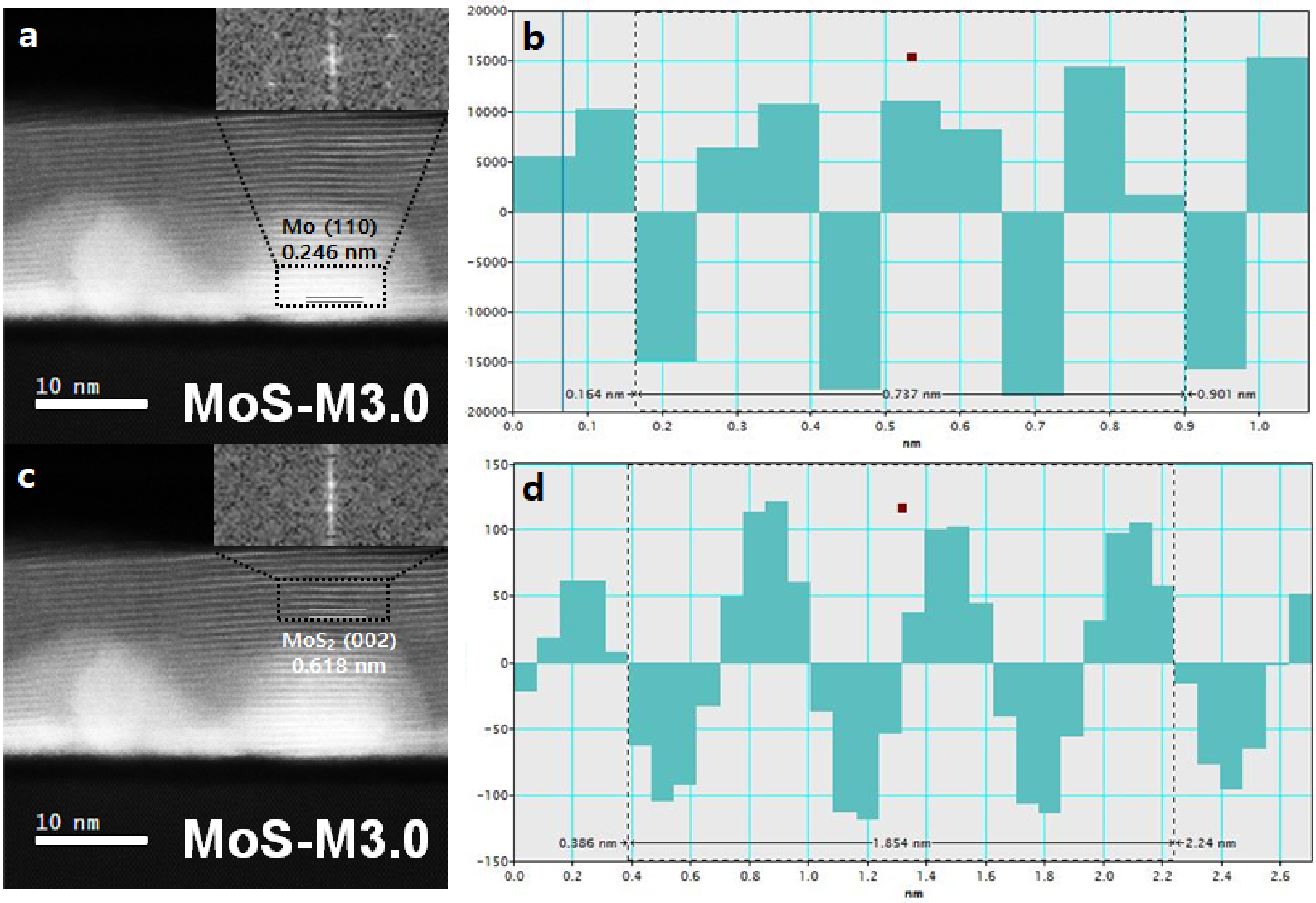

Figure S15. **a** STEM image of MoS–M3.0 with the corresponding fast Fourier transformation (FFT) pattern (inset), showing lattice fringes of the Mo (110) plane (0.246 nm). **b** Inverse fast Fourier transformation (IFFT) profile from (**a**), confirming the spacing. **c** STEM image of MoS–M3.0 with lattice fringes of the $MoS_2$ (002) plane (0.618 nm); inset shows the FFT pattern. **d** IFFT profile from (**c**), showing periodic fringes consistent with $MoS_2$ (002).

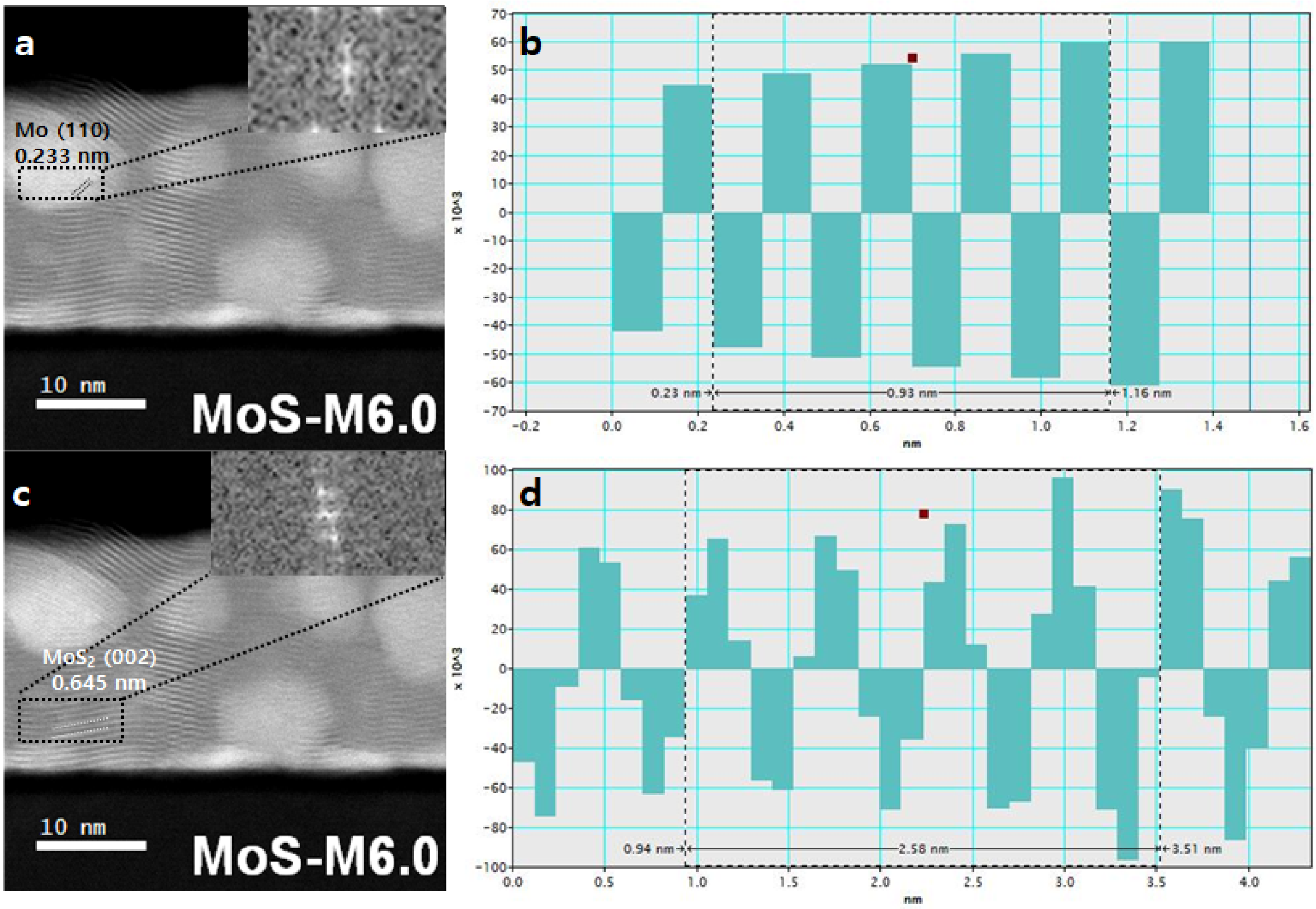

Figure S16. **a** STEM image of MoS–M6.0 with the corresponding FFT pattern (inset), showing lattice fringes of the Mo (110) plane (0.233 nm). **b** IFFT profile from (**a**), confirming the spacing. **c** STEM image of MoS–M6.0 with lattice fringes of the $MoS_2$ (002) plane (0.645 nm); inset shows the FFT pattern. **d** IFFT profile from (**c**), showing periodic fringes consistent with $MoS_2$ (002).

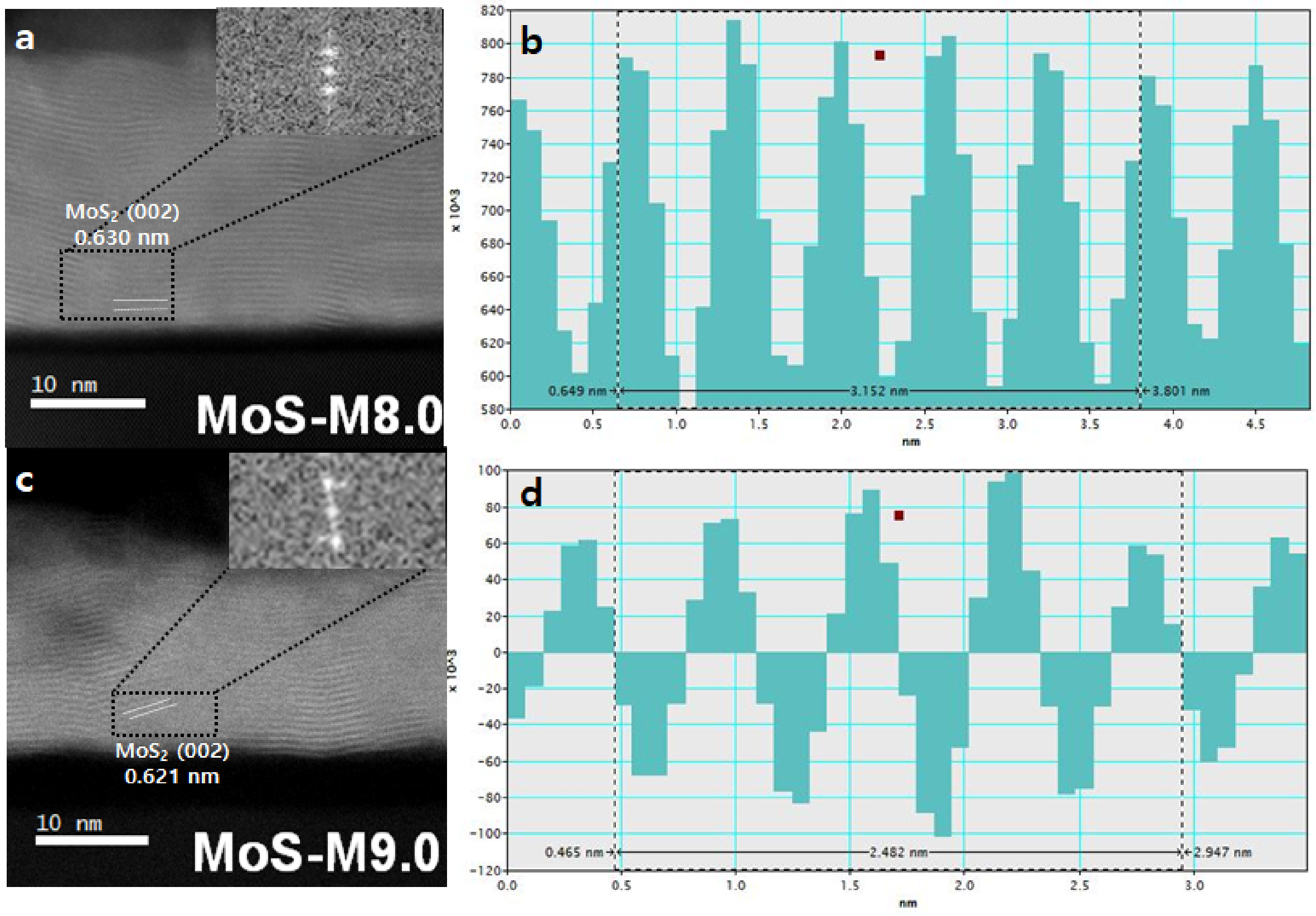

Figure S17. **a** STEM image of MoS–M8.0 with the corresponding FFT pattern (inset), showing lattice fringes of the $MoS_2$ (002) plane (0.630 nm). **b** IFFT profile from (**a**), confirming the spacing. **c** STEM image of MoS–M9.0 with lattice fringes of the $MoS_2$ (002) plane (0.621 nm); inset shows the FFT pattern. **d** IFFT profile from (**c**), showing periodic fringes consistent with $MoS_2$ (002).

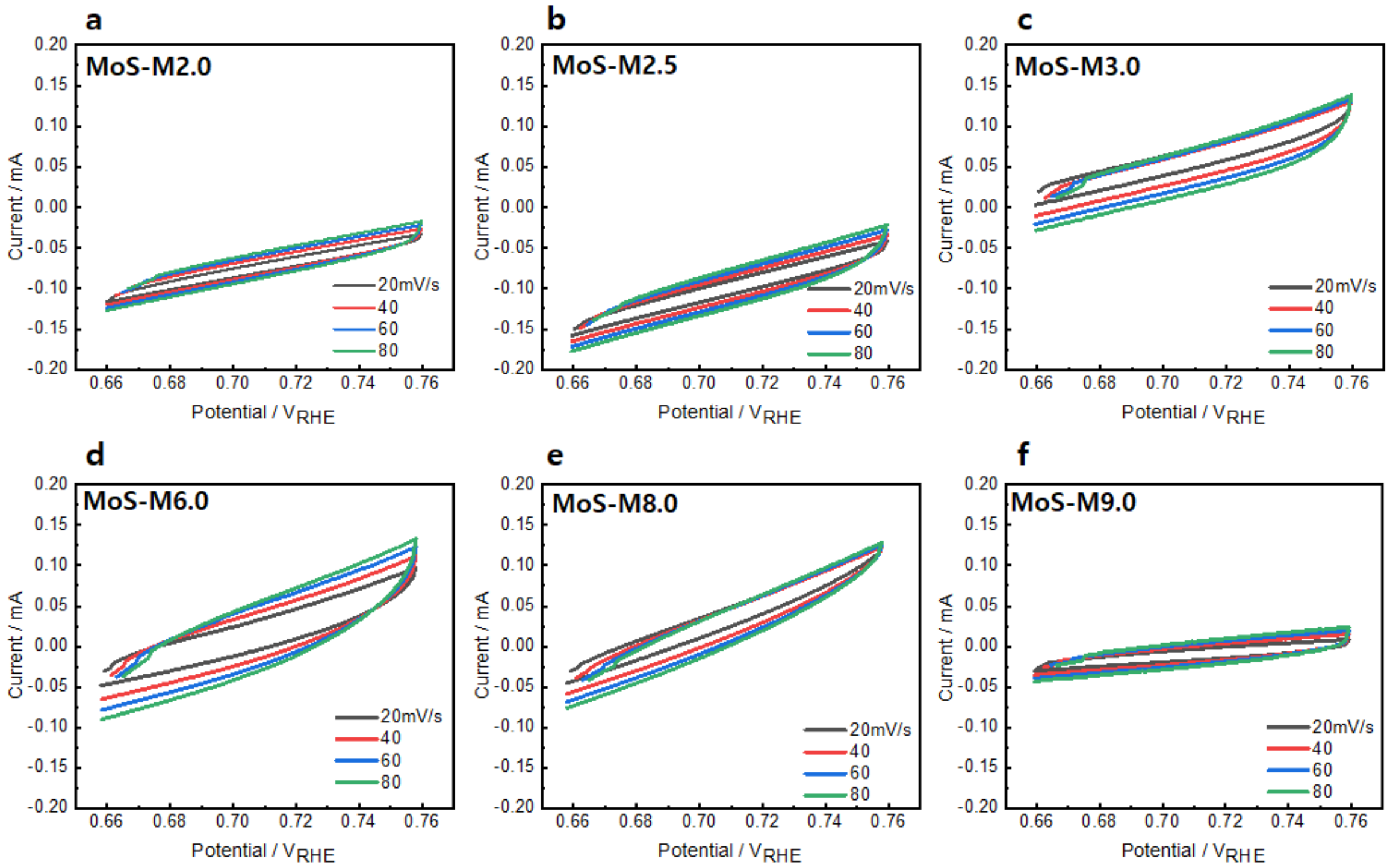

Figure S18. **a-f** Cyclic voltammetry curves of MoS-M2.0 (**a**), M2.5 (**b**), M3.0 (**c**), M6.0 (**d**), M8.0 (**e**), and M9.0 (**f**) measured at various scan rates (20, 40, 60, 80 mV $s^{-1}$) from 0.66 V to 0.76 V (vs. RHE).

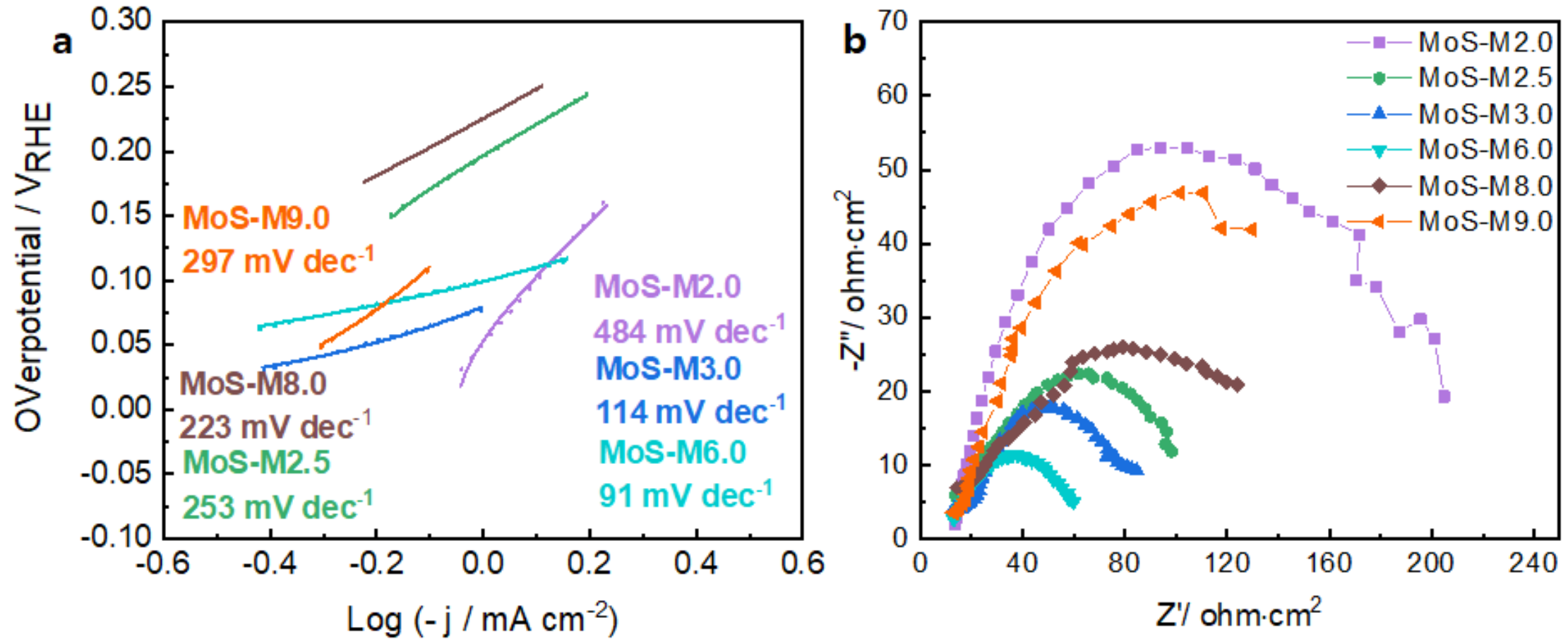

Figure S19. **a, b** Tafel plot and Nyquist plot for $MoS_2$ electrodes with different S thickness.

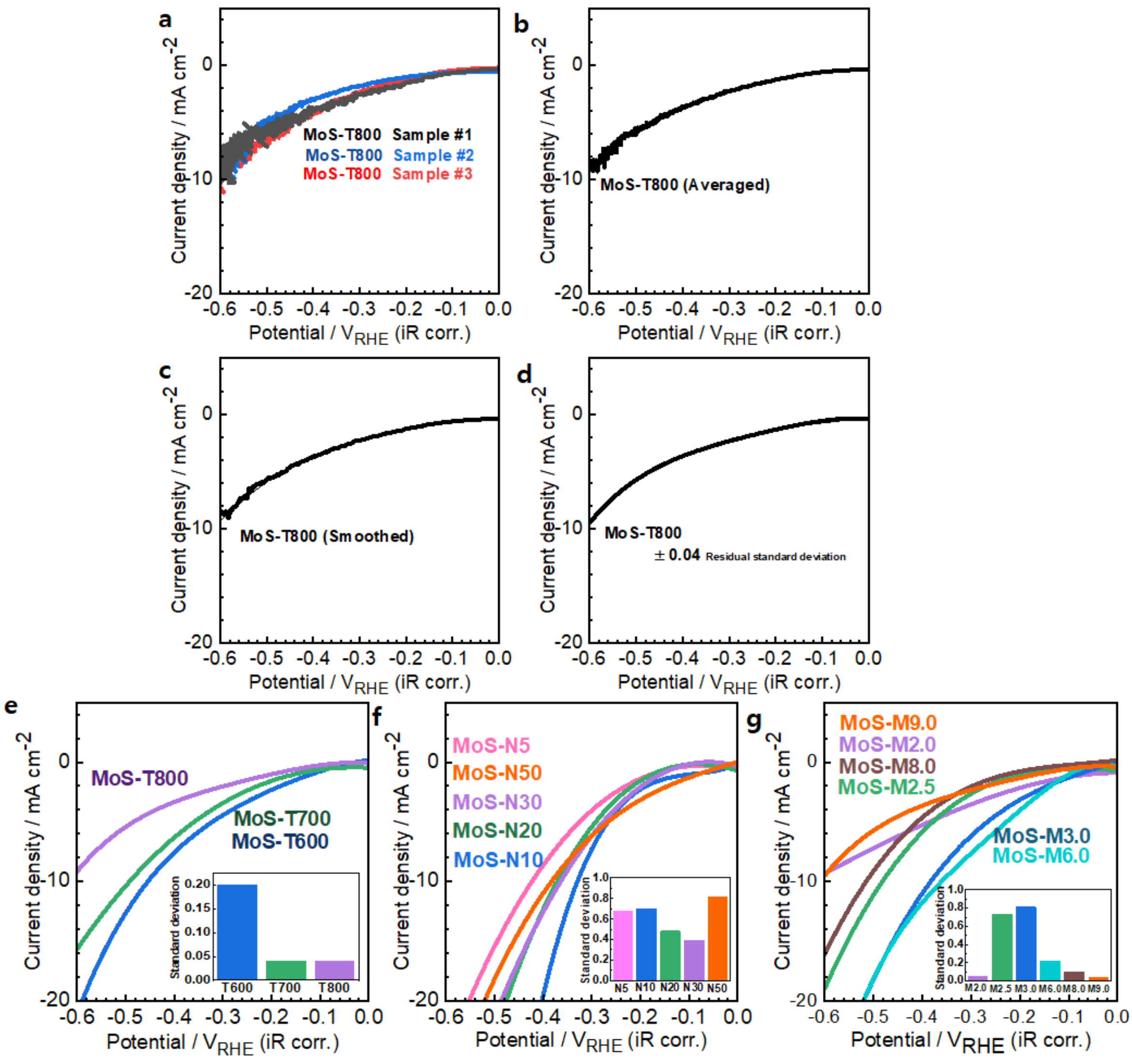

Figure S20. Data processing workflow for HER LSV polarization curves of MoS–T800: **a** Three independently measured HER LSV curves. **b** Averaged polarization curve obtained from the three measurements. **c** Smoothed LSV curve derived from the average data. **d** Trend curve with the residual standard deviation indicating the noise level of the averaged LSV. **e-g** Trend curves of $MoS_2$ films with the residual standard deviation.

Table S1. Comparison of HER performance metrics of MBE-grown $MoS_2$ with representative $MoS_2$-based catalysts reported in the literature

| Catalyst | Synthesis method | Substrate | Electrolyte | Loading amount / μg $cm^{-2}$ | Mass-based current density at -0.25$V_{RHE}$ / mA $mg^{-1}$ | $MoS_2$ mass-based TOF at -0.25$V_{RHE}$ / nmol $H_2$ $μg^{-1}s^{-1}$ | η at -10 mA $cm^{-2}$ / $V_{RHE}$ | Rct / Ohm·$cm^2$ | Tafel slope / mV $dec^{-1}$ | Ref. |
|---|---|---|---|---|---|---|---|---|---|---|
| **MoS-N10** | **Molecular beam epitaxy** | **Si wafer** | **1M KOH** | **3.7** | **1019** | **5.28** | **-0.33** | **52.8** | **80** | **This work** |
| FeP/$MoS_2$ | Hydrothermal synthesis | Glassy carbon | 0.5M $H_2SO_4$ | 1000 | 90 | 0.46 | -0.11 | 33 | 67.8 | [S1] |
| $MoS_2$/CC | Hydrothermal synthesis | Carbon cloth | 0.5M $H_2SO_4$ | 190 | 452 | 2.34 | - | - | 50 | [S2] |
| rGO/$MoS_2$-S | Hydrothermal synthesis | Glassy carbon | 1M $H_2SO_4$ | 2000 | 5 | 0.03 | -0.25 | 80 | 75 ±4 | [S3] |
| $MoS_2$/FNS/FeNi foam | Chemical Vapour Deposition | FeNi foam | 1M KOH | 150 | 565 | 2.93 | -0.12 | 4.0 | 45.1 | [S4] |
| $MoS_2$/IPC-2 | Hydrothermal synthesis | Glassy carbon | 0.5M $H_2SO_4$ | 254 | 886 | 4.59 | -0.18 | 10.1 | 38 | [S5] |
| Exfoliated 3R $MoS_2$ | Exfoliation | Glassy carbon | 0.5M $H_2SO_4$ | 106 | 94 | 0.49 | -0.25 | - | 58 | [S6] |
| $MoS_2$–$TMS_2$–$TMoO_x$ hybrids | Hydrothermal synthesis | Glassy carbon | 1M KOH | 200 | 50 | 0.26 | -0.25 | - | 74 | [S7] |
| $MoS_2$/ GO 50/1 550◦C | Impinging jet reactor. | Glassy carbon | 0.5M $H_2SO_4$ | 1131 | 19.6 | 0.10 | -0.22 | - | 84.3 | [S8[ |
| Rh-$MoS_2$ | Lithium intercalation synthesis | Glassy carbon | 0.5M $H_2SO_4$ | 309 | 13870 | 7186.04 | -0.05 | 5.1 | 24 | [S9] |
| $MoS_2$/SWNT | Exfoliation | PyC | 0.5M $H_2SO_4$ | 1450 | 5.31 | 0.03 | - | 72 | 102±17 | [S10] |